\documentclass[twocolumn,showpacs,preprintnumbers,amsmath,amssymb,prb,floatfix,superscriptaddress]{revtex4}

\usepackage{graphicx,tabularx}
\usepackage[version=3]{mhchem} 
\usepackage{amssymb}
\usepackage{dcolumn}
\usepackage{color}
\usepackage[mathcal]{euscript}
\usepackage{color}
\usepackage{subfigure}
\usepackage{multirow}
\definecolor{gray0}{gray}{0.0}
\definecolor{gray64}{gray}{0.25}
\definecolor{gray128}{gray}{0.5}
\definecolor{gray192}{gray}{0.75}
\definecolor{gray255}{gray}{1.0}

\def\Ce3{Ce$^{\mathrm{III}}$}
\def\La3{La$^{\mathrm{III}}$}
\def\Ce4{Ce$^{\mathrm{IV}}$}
\def\A3{A$^{\mathrm{III}}$}
\def\B4{B$^{\mathrm{IV}}$}

\def\CO2{CeO$_2$}
\def\C2O3{Ce$_2$O$_3$}
\def\Ef{E$_f$}

\begin{document}

\title{Aliovalent Doping of \ce{CeO2}: DFT-study of Oxidation State and Vacancy Effects.}
\author{Danny E. P. Vanpoucke}
\affiliation{Department of Inorganic and Physical Chemistry, Ghent University, Krijgslaan $281$ - S$3$, $9000$ Gent, Belgium}
\affiliation{Center for Molecular Modeling, Ghent University, Technologiepark $903$, $9052$ Zwijnaarde, Belgium}
\email[corresponding author:]{Danny.Vanpoucke@ugent.be}
\author{P. Bultinck}
\affiliation{Department of Inorganic and Physical Chemistry, Ghent University, Krijgslaan $281$ - S$3$, $9000$ Gent, Belgium}
\author{S. Cottenier}
\affiliation{Center for Molecular Modeling, Ghent University, Technologiepark $903$, $9052$ Zwijnaarde, Belgium}
\affiliation{Department of Materials Science and Engineering, Ghent University, Technologiepark $903$, $9052$ Zwijnaarde, Belgium}
\author{V. Van Speybroeck}
\affiliation{Center for Molecular Modeling, Ghent University, Technologiepark $903$, $9053$ Zwijnaarde, Belgium}
\author{I. Van Driessche}
\affiliation{Dept. Inorganic and Physical Chemistry, Ghent University, Krijgslaan $281$ - S$3$, $9000$ Gent, Belgium}

\cite{}

\date{\today}
\begin{abstract}
The modification of the properties of \ce{CeO2} through aliovalent doping are investigated within the \emph{ab initio} density functional theory framework. Lattice parameters, dopant atomic radii, bulk moduli and thermal expansion coefficients  of fluorite type \ce{Ce_{1-$x$}M_{$x$}O_{2-$y$}} (with M$ = $ \ce{Mg}, \ce{V}, \ce{Co}, \ce{Cu}, \ce{Zn}, \ce{Nb}, \ce{Ba}, \ce{La},  \ce{Sm}, \ce{Gd}, \ce{Yb}, and  \ce{Bi}) are presented for dopant concentrations in the range $0.00 \leq x \leq 0.25$. The stability of the dopants is compared and discussed, and the influence of oxygen vacancies is investigated. It is shown that oxygen vacancies tend to increase the lattice parameter, and strongly decrease the bulk modulus. Defect formation energies are correlated with calculated crystal radii and covalent radii of the dopants, but are shown to present no simple trend. The previously observed inverse relation between the thermal expansion coefficient and the bulk modulus is shown to persist independent of the inclusion of charge compensating vacancies.
\end{abstract}

\pacs{  } 
\maketitle
\section{Introduction}
\indent Cerium oxide based materials have been receiving increasing attention during the last decades. A literature search on Web of Science shows that since $2007$ about $1000$ peer reviewed journal articles a year have appeared containing ceria as a topic. The interest in these materials is due to their versatile nature in industrial applications, which generally originate from the remarkable oxidation and reduction properties of \ce{CeO2}. The major fraction of the investigations of ceria based materials are linked to solid oxide fuel cells and catalysis.\cite{TullerHL:JES1975, TrovarelliA:CatalRev1996, ManzoliMaela:CT2008, KasparJ:CatalToday1999, SteeleBCH:SolStateIon2000, ShaoZ:nature2004, YangZ:JChemPhys2007, ShapovalovV:JCatal2007, MayernickAD:JPhysChemC2008, DesaunayT:SurfSci2012, YaoHC:JCatal1984, FuQ:2003Science}
In case of the latter, these materials play both the role of catalyst support and catalyst. In addition to being used in automotive three-way-catalyst (TWC) and in water-gas-shift reactions, ceria based materials are also used as oxygen sensors, thermal barrier coatings and much more.\cite{YaoHC:JCatal1984, McBrideJR:JApplPhys1994, FuQ:2003Science, CaoXueqiang:AdvMat2003, CaoXueqiang:JEurCeramSoc2004} Recently, \ce{CeO2} and doped \ce{CeO2} have been used as buffer layers for thin film \ce{YBa2Cu3O_{7-$\delta$}} coated superconductors.\cite{ParanthamanM:1997PhysC, OhSanghyun:PhysC1998, PennemanG:2004EuroCeram, TakahashiY:PhysC2004, KnothKerstin:PhysC2005, VandeVeldeNigel:EurJInorChem2010, VyshnaviN:2012JMaterChem}\\
\indent In experiments, ceria based materials have been doped with many different types of elements.\cite{TrovarelliA:CatalRev1996, MogensenM:SolStateI2000}
These experiments also show different dopant elements to have different effects on different properties. Furthermore, based on the application of interest, dopant concentrations can vary from $<1$\% up to mixed oxides where dopant concentrations of $50$\% and more are used. In addition, also the preparation methods vary greatly (\emph{e.g.} combustion synthesis,\cite{BeraP:ChemMater2002, LiB:JPowerSources2008} chemical and physical vapor deposition,\cite{MullinsDR:SurfSci1999, AspinallHC:InorgChem2011} sol-gel deposition\cite{PennemanG:2004EuroCeram, VandeVeldeNigel:EurJInorChem2010, VyshnaviN:2012JMaterChem} \emph{etc.}), influencing the investigated properties.\cite{RossignolS:JMaterChem1999, VanDriesscheI:2002ECVII, ClaparedeL:InorChem2011, HorlaitD:InorChem2012} In contrast to all this variation, ceria based materials generally tend to have the same crystal structure (more specifically fluorite), adding to their usefulness for general applications.\\
\indent Although the body of theoretical work on ceria is smaller than the amount of experimental work published, it remains quite impressive. Much of this work focusses on a single aspect of a single application, often investigating the effect of a single dopant element.\cite{YangZ:JChemPhys2007, ShapovalovV:JCatal2007, MayernickAD:JPhysChemC2008, SongYQ:JPhysCondensMatter2009, YangZ:SurfSci2008, AnderssonDA:PNAS2006, LuZ:ChemPhysLett2011, VanpouckeDannyEP:2011PhysRevB_LCO} Investigations of series of dopant elements are much less frequent, and with only few exceptions almost exclusively focus on the lanthanide series.\cite{MinerviniL:JAmCeramSoc2000, AnderssonDA:PNAS2006, YeF:SolStateIon2009, WeiX:SolStateIonics2009} This is mainly due to the fact that this series (or elements from it) is also the most often investigated in experiments.\cite{TrovarelliA:CatalRev1996, MogensenM:SolStateI2000, McBrideJR:JApplPhys1994, SteeleBCH:SolStateIon2000, RyanKM:JPCondMat2003, LopesFWB:Hydrometallurgy2009, ReddyB:2010CM, HorlaitD:2011InorChem, ClaparedeL:InorChem2011, VyshnaviN:2012JMaterChem, HorlaitD:InorChem2012} Recently, also the series of tetravalent/group IV elements have been investigated by means of \emph{ab initio} calculations. Andersson \emph{et al.} focused on the ionic conductivity properties of oxygen vacancies for tetravalent dopants in \ce{CeO2}.\cite{AnderssonDA:2007bPhysRevB, AnderssonDA:2007ApplPhysLett} The present authors investigated the stability and influence of group IV dopants on mechanical and structural properties of \ce{CeO2}.\cite{VanpouckeDannyEP:2012aApplSurfSci, VanpouckeDanny:2012dGroupIVdopants}\\
\indent With the large variety of applications comes a large variation of the desired properties. For example, for a system to be a good solid oxide fuel cell, it should exhibit high ionic conductivity, whereas for it to be a good buffer layer, it should have a low ionic conductivity.
As a second example, lattice matching through doping of a buffer layer requires a homogenous distribution of the dopants in the bulk of the material, while doping of a catalyst often benefits from dopants residing at or near the surface.\\
\indent Because of this, we refrain from focussing on one specific application in this work, and present general trends instead. This paper expands on our previous work through the study of aliovalent dopants, and the introduction of charge compensating vacancies.\cite{VanpouckeDanny:2012dGroupIVdopants, fn:Alio:CompChargeVacDef} For practical reasons we have limited our work to a subset of the aliovalent dopants investigated in experimental work: \ce{Mg}, \ce{V}, \ce{Co}, \ce{Cu}, \ce{Zn}, \ce{Nb}, \ce{Ba}, \ce{La}, \ce{Sm}, \ce{Gd}, \ce{Yb}, and  \ce{Bi}.\cite{TrovarelliA:CatalRev1996,KundakovicLj:ApplCatalA1998, KundakovicLj:JCat1998, LiB:JPowerSources2008, ManzoliMaela:CT2008, McBrideJR:JApplPhys1994, BeraP:ChemMater2002, MogensenM:SolStateI2000, WangX:JPhysChemB2005, WangX:JPhysChemB2006, KnothKerstin:PhysC2005, deBiasi:JSolStateChem2005, LiB:IJHE2010, VandeVeldeNigel:EurJInorChem2010, AnwarMS:JAlloysCompd2011, VyshnaviN:2012JMaterChem, BrisseF:CJChem1967, RyanKM:JPCondMat2003, BaeJongSung:JoECS2004, deBiasi:JAlloysCompd2008, SheYusheng:IJHE2009, AiniradA:JAlloysCompd2011, TiwariA:ApplPhysLett2006,VodungboB:ApplyPhysLett2007, FernandesV:PhysRevB2007, SongYQ:JApplPhys2007, WenQY:JPhysCondensMatter2007, SinghalRK:JPhysDApplPhys2011, SacanellJ:ApplPhysLett2012, XuD:SolStateIonics2011, HorlaitD:2011InorChem, YaoHC:JPowerSources2012}\\
\indent In this paper, we investigate the influence of aliovalent doping on the properties of \ce{CeO2} using \textit{ab-initio} density functional theory (DFT) calculations. The theoretical methods and different supercells used are presented in Sec.~\ref{CeO2pX:sc_theormeth}. To study the contributions due to the dopants and vacancies separately, we first considering systems containing dopants only (Sec.~\ref{CeO2pX:sc_Result_AlioNV}), and then systems containing combinations of dopants and charge compensating oxygen vacancies (Sec.~\ref{CeO2pX:sc_IncludeVac}). For systems without oxygen vacancies, the atomic radii of the dopants are calculated and compared to the values tabulated for the Shannon atomic crystal radii.\cite{Shannon:table, Shannon:ACSA1976} Concentration dependent defect formation energies are calculated and put in relation with the calculated dopant radii and covalent dopant radii. The change in the bulk modulus (BM) and thermal expansion coefficient (TEC) of \ce{CeO2} due to dopants  is studied, and is shown to follow opposing trends. In addition, we investigate the modification, due to charge compensating vacancies, of the dopant (Cu, Zn, and Gd) influence on the BM, defect formation energy and lattice parameter. Summary and conclusions are presented in Sec.~\ref{CeO2pX:sc_Conclusion}.
\begin{figure}[tb]
\begin{center}
    \includegraphics[width=8cm,keepaspectratio=true]{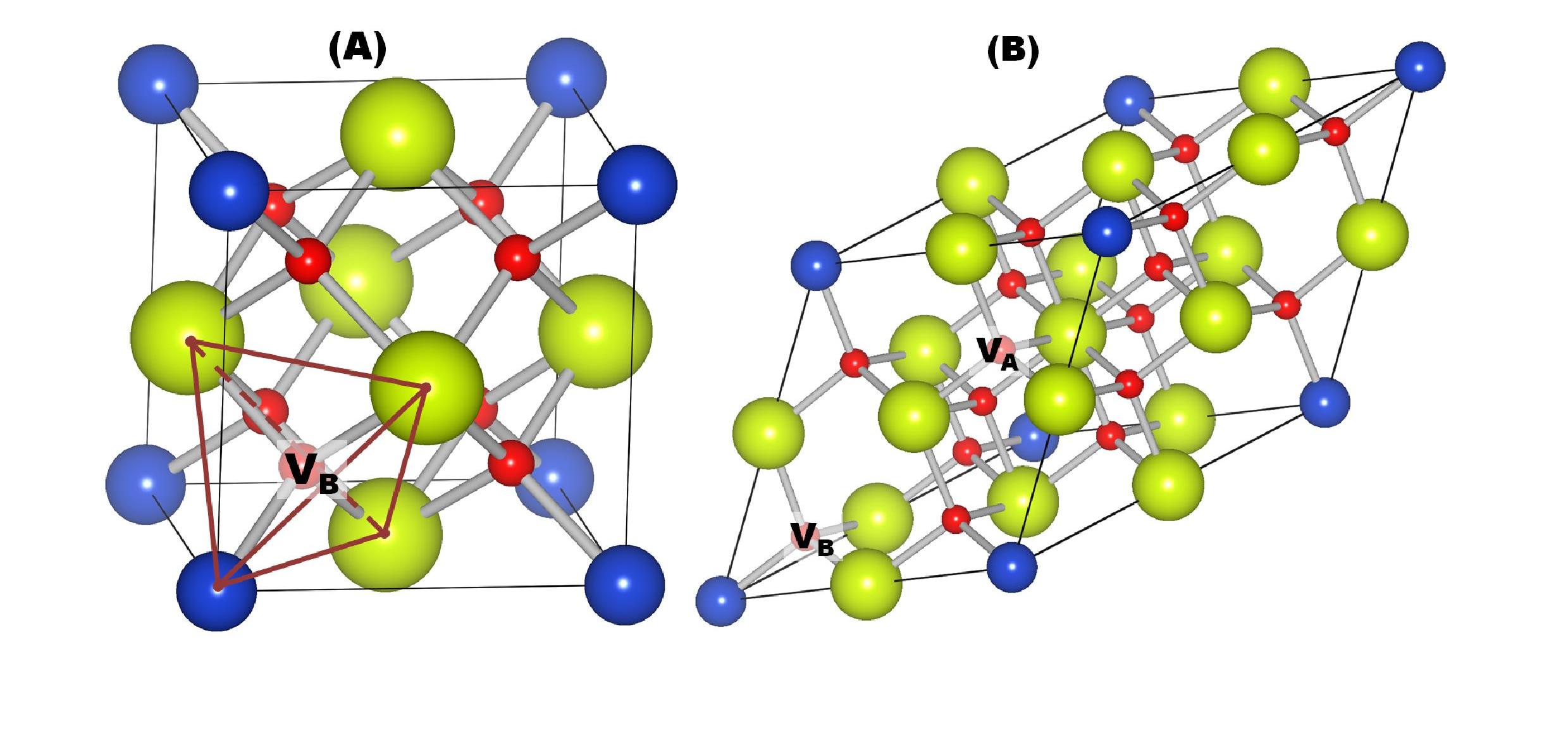}\\
\end{center}
  \caption[Ball-and-stick representations of doped \ce{CeO2} supercells]{(color online) Ball-and-stick representations of doped \ce{CeO2} $\mathrm{c}111$ (a) and $\mathrm{p}222$ (b) supercells. Yellow (red) spheres indicate the positions of the Ce (O) atoms, while the dopant position is given by the blue sphere. Vacancy positions are indicated (V$_\mathrm{A}$ and V$_\mathrm{B}$), as is the surrounding tetrahedron (red lines). The single dopant/oxygen vacancy gives rise to a dopant/vacancy concentration of $25$\%/$12.5$\% in the $\mathrm{c}111$, and $12.5$\%/$6.25$\% in the $\mathrm{p}222$ supercell, respectively.}\label{fig:c111p222_geom}
\end{figure}
\begin{figure}[tb]
\begin{center}
    \includegraphics[width=8cm,keepaspectratio=true]{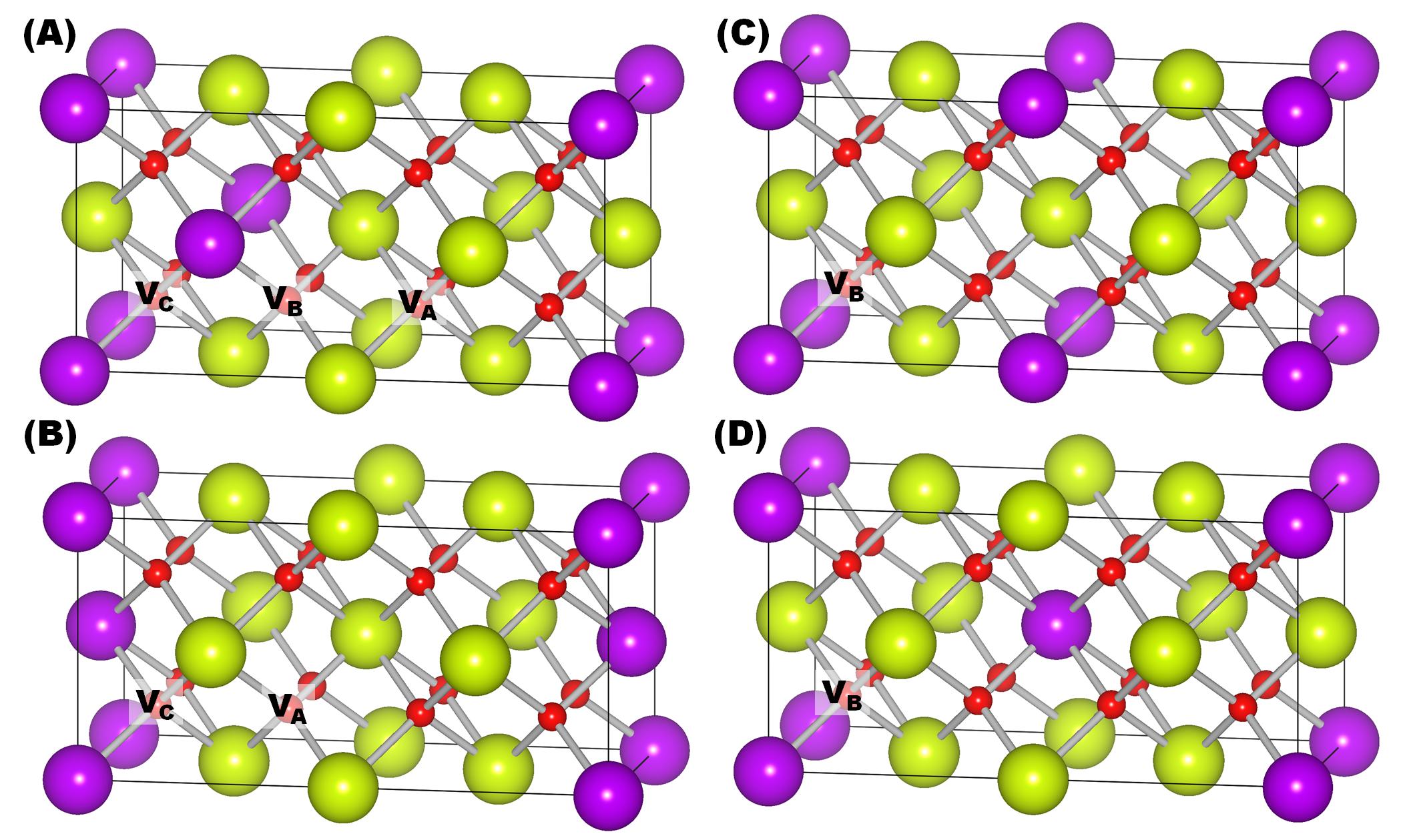}\\
\end{center}
  \caption[Ball-and-stick representations of \ce{Ce_{0.75}Gd_{0.25}O_{1.875}} configurations]{(color online) Ball-and-stick representations of different \ce{Ce_{0.75}Gd_{0.25}O_{1.875}} configurations in a double c$111$ supercell. Yellow, red, and purple spheres indicate the positions of the Ce, O, and Gd atoms. Possible vacancy positions are indicated (V$_\mathrm{A}$, V$_\mathrm{B}$, and V$_\mathrm{C}$).}\label{fig:c111A1x_Gd_geom}
\end{figure}
\section{Computational setup}\label{CeO2pX:sc_theormeth}
\indent We perform \textit{ab-initio} density functional theory (DFT) calculations using the projector augmented waves (PAW) method as implemented in the Vienna \textit{ab-initio} Package (\textsc{vasp}) program. The LDA functional as parameterized by Ceperley and Alder and the GGA functional as constructed by Perdew, Burke and Ernzerhof (PBE) are used to model the exchange and correlation behavior of the electrons.\cite{Blochl:prb94, Kresse:prb99, CA:prl1980, PBE_1996prl, Kresse:prb93, Kresse:prb96, fn:missLDA} From previous work it is clear that, for this type of system, the obtained results give the same qualitative picture as results obtained within the DFT+U framework.\cite{VanpouckeDannyEP:2011PhysRevB_LCO, VanpouckeDannyEP:2012aApplSurfSci, VanpouckeDanny:2012dGroupIVdopants} The plane wave kinetic energy cutoff is set to $500$ eV.\\
\indent To optimize the structures, a conjugate gradient method is used. During relaxation both atom positions and cell-geometry are allowed to change simultaneously. The convergence criterion is set to the difference in energy between subsequent steps becoming smaller than $1.0\times10^{-6}$ eV.\\
\indent The TEC are calculated as the numerical derivative of V(T) data. These V(T) data in turn are obtained through minimization of the thermal non-equilibrium Gibbs function, which is calculated using the quasi-harmonic Debye approximation,\cite{Maradudin:TheoryLattDynHarmApprox1971, BlancoMA:JMolStruc1996, FranciscoE:PRB2001} and is implemented as a module in our in-house developed \textsc{hive} code.\cite{HIVE_REFERENCE}
The BM is calculated by fitting E(V) data from fixed volume calculations to the third order isothermal Birch-Murnaghan equation of state.\cite{MurnaghanFD:PNAS1944, BirchF:PhysRev1947}
\setcounter{paragraph}{0}
\paragraph{Non-vacancy systems:\\}
\indent Symmetric supercells, containing a single dopant per supercell are used to simulate homogeneous distributions of the dopants without charge compensating vacancies.\cite{fn:Alio:CompChargeVacDef} For all these systems, relaxations started from the fluorite geometry (space group $Fm\bar{3}m$), while maintaining the crystal symmetry. The supercells used are the fluorite cubic $1\times 1\times 1$ cell with $12$ atoms (c$111$), the primitive $2\times 2\times 2$ cell with $24$ atoms (p$222$), the primitive $3\times 3\times 3$ cell with $81$ atoms (p$333$) and the cubic $2\times 2\times2$ cell with $96$ atoms (c$222$). Replacing a single Ce atom with a dopant element results in dopant concentrations of $25, 12.5, 3.7037,$ and $3.125$ \%, respectively. The doped c$111$ and p$222$ supercells are shown in Fig.~\ref{fig:c111p222_geom}, with the dopant element position indicated by the blue spheres.\\
\indent Monkhorst-Pack special $k$-point grids are used to sample the Brillouin zone.\cite{Monkhorst:prb76} For the two smaller cells we use an $8\times 8\times8$ $k$-point grid while for the two large supercells a $4\times 4\times4$ $k$-point grid is used.
\paragraph{Systems containing compensating O vacancies:\newline}
\indent For doped systems containing a single oxygen vacancy, only the c$111$ and p$222$ supercells are used, giving rise to dopant concentrations of $25$ and $12.5$\%, respectively, and oxygen vacancy concentrations of $12.5$ and $6.25$\%, respectively. The c$111$ and p$222$ configurations are shown in Fig.~\ref{fig:c111p222_geom}, where possible oxygen vacancy positions are labeled V$_\mathrm{A}$ and V$_\mathrm{B}$ (\textit{cf.} further). Every oxygen atom is positioned at the center of a cation-tetrahedron, as is shown in Fig.~\ref{fig:c111p222_geom}. As a result every vacancy site can have up to four dopant atoms as nearest neighbor. For calculations containing two dopants and one vacancy a double c$111$ supercell is used. Four inequivalent dopant distributions (A, B, C, and D) are used, shown in Fig.~\ref{fig:c111A1x_Gd_geom}. For the investigation of the influence of oxygen vacancies, we assume a homogeneous distribution of the vacancies,\cite{fn:lowConc} just as we did for the dopants. Effects due to clustering are beyond the scope of this work, and as such will not be treated.\\
\indent In this work vacancy sites with $0$ neighboring dopants are indicated as V$_{\mathrm{A}}$, if $1$ dopant is present in the surrounding tetrahedron it is referred to as V$_{\mathrm{B}}$, and if $2$ dopants are present as V$_{\mathrm{C}}$ (\textit{cf.}~Figs.~\ref{fig:c111p222_geom} and \ref{fig:c111A1x_Gd_geom}). Since only single oxygen vacancies are present, all Ce atoms in each of the systems will either be $7$-or $8$-coordinated.\\
\indent Similar  as for the supercells without vacancies, Monkhorst-Pack special $k-$point grids of $8\times 8\times 8$ grid points are used for the c$111$ and p$222$ cells.\cite{Monkhorst:prb76} For the double c$111$ supercells a $4\times 8\times 8$ grid is used instead.
\section{Aliovalent dopants without compensating oxygen vacancies}\label{CeO2pX:sc_Result_AlioNV}
\indent The use of aliovalent dopants in \ce{CeO2} introduces two (related) complications from the theoretical point of view. Firstly, aliovalent dopants give rise to charge compensating vacancies, which increases the number of possible configurations per dopant concentration significantly if the ground state configuration is not known. Secondly, since many elements can have multiple oxidation states this introduces additional uncertainties with regard to the number of required compensating vacancies and thus the ground state crystal structure.\\
\indent For these reasons, we start by investigating uncompensated dopants in fluorite \ce{Ce_{1-$x$}M_{$x$}O2} with M=Mg, V, Co, Cu, Zn, Nb, Ba, La, and Bi. This has the advantage that only effects directly due to the aliovalent dopants are observed. In Sec.~\ref{CeO2pX:sc_IncludeVac} compensating vacancies are added. This allows us to discriminate between dopant and oxygen vacancy induced changes of the investigated property. In addition, the uncompensated situation can be physically interpreted as doped systems under highly oxidizing atmosphere, which may be of interest for catalytic processes in for example automotive TWC.\cite{YaoHC:JCatal1984, TrovarelliA:CatalRev1996, KasparJ:CatalToday1999, DeganelloF:SSI2002, DeganelloF:SSI2003}
\subsection{Dopant radii and Vegard's law}
\indent In previous work, we have shown that for cubic systems without oxygen vacancies the radius of the dopant element can be calculated as:\cite{VanpouckeDannyEP:2012aApplSurfSci, VanpouckeDanny:2012dGroupIVdopants}
\begin{equation}
\mathrm{R}_M=\Big(\frac{\sqrt{3}}{4}a_{Ce_{1-x}M_xO_2} - \mathrm{R}_O - (1-n_x)\mathrm{R}_{Ce} \Big)/n_x,\label{CeO2pX:eq:Rm_def}
\end{equation}
with $n_x$ the dopant concentration, $a_{Ce_{1-x}M_xO_2}$ the lattice parameter of the doped system, and R$_O$ and R$_{Ce}$ the radii of O and Ce, respectively. From this the empirical Vegard law was obtained.\cite{VegardsLaw_DentonAR:PhysRevA1991, VanpouckeDannyEP:2012aApplSurfSci} In doping experiments, lattice parameters are often linearly fitted with regard to the dopant concentration. Deviation with respect to this Vegard law behavior is interpreted as being due to the presence of secondary phases, phase transitions or saturation, depending on the observed deviation.\cite{MorrisBC:JMatChem1993, RyanKM:JPCondMat2003, BaeJongSung:JoECS2004, BelliereV:JPhysChemB2006, WenQY:JPhysCondensMatter2007, HorlaitD:2011InorChem, VyshnaviN:2012JMaterChem}

\begin{figure}[t]
\begin{center}
    \includegraphics[width=8cm,keepaspectratio=true]{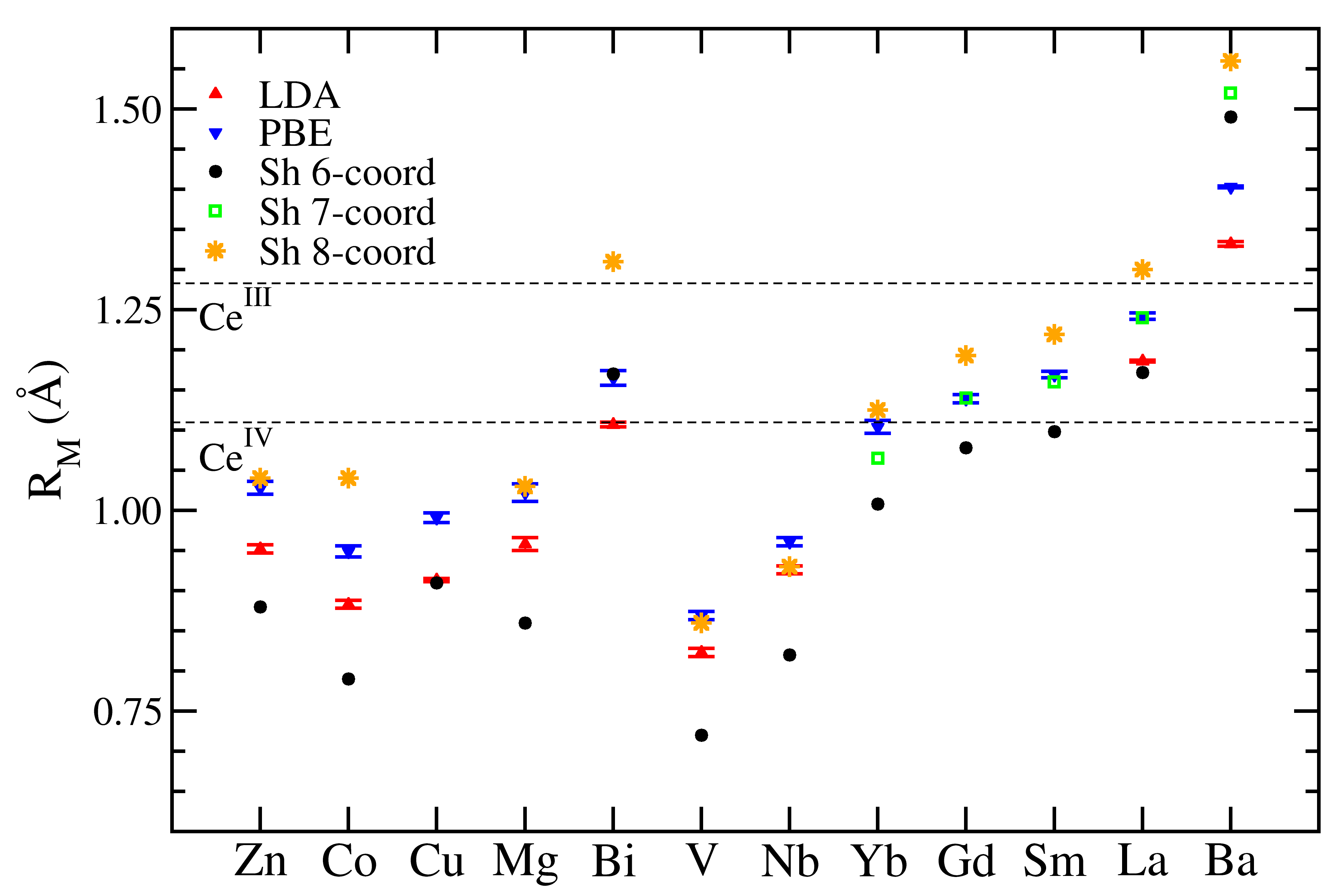}\\
\end{center}
  \caption[Calculated dopant radii for aliovalent dopants]{Comparison of calculated dopant radii in \ce{Ce_{1-$x$}M_{$x$}O2} to the Shannon crystal radius for M=\ce{Mg^{II}}, \ce{V^{IV}}, \ce{Co^{II}}, \ce{Cu^{I}}, \ce{Zn^{II}}, \ce{Nb^{IV}}, \ce{Ba^{II}}, \ce{La^{III}}, \ce{Sm^{III}}, \ce{Gd^{III}}, \ce{Yb^{III}}, and \ce{Bi^{III}} with coordination numbers $6$, $7$, and $8$ (where available).\cite{Shannon:table, Shannon:ACSA1976} The standard deviation is shown as error bars. The dopant elements are sorted according their Covalent radius, with Zn the smallest and Ba the largest element.\cite{Cordero:DT2008} The Shannon crystal radii for $8$-coordinate \ce{Ce^{III}} and \ce{Ce^{IV}} are indicated with dashed lines.\cite{Shannon:table, Shannon:ACSA1976}}\label{fig:Rm_vs_Sh}
\end{figure}

\begin{table*}[!tb]
\caption[Dopant radii and Vegard law for aliovalent dopants]{Dopant radii calculated using Eq.~(\ref{CeO2pX:eq:Rm_def}), averaged over the four dopant concentrations (avg), and standard deviation (stdev) of this value. This is done for both LDA and PBE calculated geometries. $a$ and $b$ are the intercept and slope of Vegard's law linear fit (\emph{cf.}~Eq.~\eqref{eq:VegardLinFit}) to the calculated geometries for doped CeO$_2$ systems. Lattice parameters at zero Kelvin $a_0$ and room temperature (RT) $a_{\mathrm{RT}}$ ($300$ K) are given for \ce{Ce_{0.75}M_{0.25}O2}. The \ce{CeO2} values are given as reference.\cite{fn:missLDA}} \label{table:MetalX_Rm_Vegards_aRT}
\begin{ruledtabular}
\begin{tabular}{l|cccc|crcr|cccc}
 & \multicolumn{4}{c|}{R$_M$ (\AA)} & \multicolumn{4}{c|}{Vegard's Law} & \multicolumn{2}{c}{LDA} & \multicolumn{2}{c}{PBE}\\
 & \multicolumn{2}{c}{LDA} & \multicolumn{2}{c|}{PBE} & \multicolumn{2}{c}{LDA} & \multicolumn{2}{c|}{PBE} & $a_0$ & $a_{\mathrm{RT}}$ & $a_0$ & $a_{\mathrm{RT}}$\\
 & avg & stdev & avg & stdev & $a$(\AA) & \multicolumn{1}{c}{$b$} & $a$(\AA) & \multicolumn{1}{c|}{$b$} & (\AA) & (\AA) & (\AA) & (\AA) \\
\hline
& & & & & & & & & & & & \\[-2mm]
CeO$_2$ & $1.0819$$^a$ & $0.0001$ & $1.1257$$^a$ & $0.0004$ & \multicolumn{4}{c|}{} & $5.362$ & $5.388$ & $5.463$ & $5.492$ \\
\hline
& & & & & & & & & & & & \\[-2mm]
Mg   & $0.958$ & $0.008$ & $1.022$ & $0.011$ & $5.364$ & $-0.315$ & $5.465$ & $-0.273$ & $5.285$ & $5.316$ & $5.396$ & $5.432$ \\
V    & $0.823$ & $0.005$ & $0.870$ & $0.005$ & $5.363$ & $-0.613$ & $5.464$ & $-0.610$ & $5.209$ & $5.235$ & $5.312$ & $5.341$ \\
Co   & $0.883$ & $0.005$ & $0.949$ & $0.007$ & $5.363$ & $-0.478$ & $5.464$ & $-0.427$ & $5.243$ & $5.273$ & $5.357$ & $5.392$ \\
Cu   & $0.913$ & $0.002$ & $0.991$ & $0.006$ & $5.362$ & $-0.395$ & $5.463$ & $-0.307$ & $5.264$ & $5.299$ & $5.387$ & $5.428$ \\
Zn   & $0.952$ & $0.005$ & $1.028$ & $0.008$ & $5.363$ & $-0.317$ & $5.464$ & $-0.239$ & $5.283$ & $5.315$ & $5.404$ & $5.440$ \\
Nb   & $0.926$ & $0.005$ & $0.961$ & $0.005$ & $5.363$ & $-0.375$ & $5.464$ & $-0.395$ & $5.269$ & $5.292$ & $5.365$ & $5.392$ \\
Ba   & $1.332$ & $0.003$ & $1.403$ & $0.001$ & $5.363$ & $0.566$ & $5.464$ & $0.635$ & $5.504$ & $5.533$ & $5.622$ & $5.656$ \\
La  & $1.186$ & $0.001$ & $1.242$ & $0.004$ & $5.362$ & $0.237$ & $5.464$ & $0.260$ & $5.422$ & $5.448$ & $5.529$ & $5.559$ \\
Sm   & $-$ & $-$ & $1.169$ & $0.004$ & $-$ & $-$ & $5.464$ & $0.095$ & $-$ & $-$ & $5.487$ & $5.517$ \\
Gd   & $-$ & $-$ & $1.139$ & $0.005$ & $-$ & $-$ & $5.464$ & $0.015$ & $-$ & $-$ & $5.468$ & $5.498$ \\
Yb   & $-$ & $-$ & $1.104$ & $0.008$ & $-$ & $-$ & $5.464$ & $-0.056$ & $-$ & $-$ & $5.449$ & $5.482$ \\
Bi   & $1.107$ & $0.003$ & $1.165$ & $0.009$ & $5.363$ & $0.044$ & $5.465$ & $0.060$ & $5.373$ & $5.400$ & $5.480$ & $5.511$ \\
\end{tabular}
\end{ruledtabular}
\begin{flushleft}
$^a$ The Ce radius is calculated using Eq.~(\ref{CeO2pX:eq:Rm_def}), where the $4$-coordinated Shannon crystal radius for oxygen is taken as $1.24$ \AA\ \cite{Shannon:table, Shannon:ACSA1976}.\\
\end{flushleft}
\end{table*}
\indent Table~\ref{table:MetalX_Rm_Vegards_aRT} shows the calculated dopant radii and coefficients of Vegard's Law. The intercept $a$ and slope $b$ of the latter are found from rewriting Eq.~\eqref{CeO2pX:eq:Rm_def} as
\begin{equation}\label{eq:VegardLinFit}
a_{Ce_{1-x}M_xO_2}=a_{CeO_2} + \left(\frac{4}{\sqrt{3}}(R_O+R_M)-a_{CeO_2}\right)n_x,
\end{equation}
as was shown in previous work.\cite{VanpouckeDannyEP:2012aApplSurfSci} The small standard deviations on the calculated dopant radii ($\leq0.01$\AA) show consistent values are found for the systems of different concentrations. The calculated lattice parameter for \ce{Ce_{0.75}Sm_{0.25}O_{2}} seems to be in good agreement with the experimental lattice parameter of $5.4314$\AA\ for \ce{Ce_{0.8}Sm_{0.2}O_{2-$\delta$}} by Yao \textit{et al.}\cite{YaoHC:JPowerSources2012} and $\sim5.435$\AA\ for \ce{Ce_{0.85}Sm_{0.15}O_{1.925}} by Xu \textit{et al.}\cite{XuD:SolStateIonics2011}, knowing that PBE generally overestimates lattice parameters by a few percent. Also the very small variation of the experimental lattice parameter with the Sm concentration is in agreement with the calculated slope of the Vegard law, if one takes into account that the different synthesis methods have an influence on the obtained lattice parameters.\cite{XuD:SolStateIonics2011, YaoHC:JPowerSources2012} Yao \emph{et al.} also calculated the Vegard law slope for Co doped \ce{Ce_{0.8}Sm_{0.2}O_{2-$\delta$}} and find a lattice contraction, in qualitative agreement with our theoretical results.\cite{YaoHC:JPowerSources2012} The smaller experimental lattice contraction is mainly due to the presence of oxygen vacancies, which, as will be shown later for other aliovalent dopants, gives rise to a lattice expansion compared to a system without oxygen vacancies, and thus compensates the lattice contraction due to the Co dopants to some extent, lowering the degree of lattice contraction.\\
\indent In Fig.~\ref{fig:Rm_vs_Sh} the calculated atomic radii are compared to the Shannon crystal radii for $6$-,$7$-, and $8$-coordinated configurations.\cite{Shannon:table, Shannon:ACSA1976} For tetravalent V and Nb, both LDA and PBE results are in good agreement with the $8$-coordinate Shannon crystal radius, while for trivalent Yb, Gd, Sm, and La the $7$-coordinate values give the best agreement, despite the fact that all dopants are placed in the $8$-coordinate environment of \ce{CeO2}.\cite{fn:ValenceDisclaimer} Furthermore, the radii for divalent Mg and Zn show good agreement with the $8$-coordinate radii, while monovalent Cu and trivalent Bi present $6$-coordinate radii. The deduced $6$-coordination for Cu shows nice agreement with the coordination number $5$--$6$ obtained by Wang \emph{et al.} from X-ray adsorption fine structure (XAFS) measurements.\cite{WangX:JPhysChemB2005} It differs, however, from the $4$-coordination found in calculations by Lu \textit{et al.}\cite{LuZ:ChemPhysLett2011} where a broken symmetry structure for the Cu doped \ce{CeO2} was used. The resulting tetragonal structure for such a broken symmetry system is $0.667$\% larger in volume than the cubic fluorite structure used in this work, making the calculated atomic radius for Cu slightly larger than the one presented. The value for divalent Co in turn tends toward $7$-fold coordination. Note that the Shannon crystal radii for \ce{Co^{$\mathrm{III}$}} and \ce{Co^{$\mathrm{IV}$}} would be too small,\cite{Shannon:table, Shannon:ACSA1976} showing that the divalent nature, inferred from the calculated radius, supports the experimental suggestion of divalent Co dopants.\cite{TiwariA:ApplPhysLett2006, SongYQ:JApplPhys2007, WenQY:JPhysCondensMatter2007, SinghalRK:JPhysDApplPhys2011, fn:magCo}\\
\indent The results for Ba are a bit peculiar, since the calculated radius is significantly lower than either $6$-, $7$-, or $8$-coordinate Shannon crystal radii for divalent Ba.\cite{Shannon:table, Shannon:ACSA1976} Assuming the general trends seen in the Shannon crystal radii for other elements are also valid for Ba (i.e. increasing valence results in decreasing radius under constant coordination) this would lead to the conclusion that Ba behaves as having an oxidation state higher than $\mathrm{II}$ when used as a dopant for \ce{CeO2}, which is puzzling.\\
\indent In conclusion, in contrast to our previous work on group IV elements, aliovalent dopants tend not to present full $8$-coordination, but rather act as if they are only $7$- or $6$-coordinated.\cite{VanpouckeDannyEP:2012aApplSurfSci, VanpouckeDanny:2012dGroupIVdopants}\\
\indent On the other hand, as might be expected, perfect Vegard law behavior is observed for all the systems under investigation. Combined with the calculated atomic radii, this provides a way to experimentally estimate the valence of dopant elements based on the obtained lattice parameter under oxidizing atmosphere. This is done by calculating the atomic crystal radius of the dopant based on the measured lattice parameter, and then comparing this radius to the tabulated values by Shannon,\cite{Shannon:table, Shannon:ACSA1976} to deduce the dopant valence.
\begin{table}[!bt]
\caption[Defect formation energies for aliovalent dopants]{Defect formation energy \Ef\ for doped \ce{CeO2} at different dopant concentrations.\cite{fn:missLDA}}\label{table:MetalXsubst_energies}
\begin{ruledtabular}
\begin{tabular}{l|rrrr}
 & \multicolumn{4}{c}{\Ef (eV)} \\
 &  25\% & 12.5\% & 3.704\% & 3.125\% \\
\hline
 & \multicolumn{4}{c}{LDA}  \\
\hline
CeO$_2$ & \multicolumn{4}{c}{-11.484$^{a}$} \\
Mg & $8.221$ & $8.409$ & $8.470$ & $8.475$   \\
V  & $6.243$ & $6.313$ & $6.338$ & $6.322$   \\
Co & $12.353$ & $12.425$ & $12.479$ & $12.473$   \\
Cu & $13.517$ & $13.464$ & $13.463$ & $13.458$   \\
Zn & $11.465$ & $11.651$ & $11.696$ & $11.707$   \\
Nb & $3.738$ & $3.445$ & $3.400$ & $3.415$   \\
Ba & $7.621$ & $7.777$ & $7.933$ & $7.956$   \\
La & $2.403$ & $2.389$ & $2.418$ & $2.422$   \\
Bi & $7.902$ & $8.069$ & $8.114$ & $8.095$  \\
\hline
 & \multicolumn{4}{c}{PBE} \\
\hline
CeO$_2$ & \multicolumn{4}{c}{-10.418$^{a}$}  \\
Mg & $8.036$ & $8.223$ & $8.275$ & $8.284$   \\
V  & $6.256$ & $6.320$ & $6.348$ & $6.361$   \\
Co & $11.750$ & $11.780$ & $11.800$ & $11.801$   \\
Cu & $12.922$ & $12.878$ & $12.878$ & $12.879$   \\
Zn & $11.057$ & $11.249$ & $11.282$ & $11.300$   \\
Nb & $4.059$ & $3.746$ & $3.726$ & $3.761$   \\
Ba & $7.518$ & $7.681$ & $7.850$ & $7.882$   \\
La & $2.438$ & $2.429$ & $2.464$ & $2.469$   \\
Sm & $-2.181$ & $-2.218$ & $-2.228$ & $-2.236$   \\
Gd & $2.396$ & $2.445$ & $2.449$ & $2.448$   \\
Yb & $4.438$ & $4.495$ & $4.508$ & $4.500$   \\
Bi & $7.912$ & $8.069$ & $8.111$ & $8.093$  \\
\end{tabular}
\end{ruledtabular}
\begin{flushleft}
$^{a}$ Instead of the defect formation energy the heat of formation is given.\\
\end{flushleft}
\end{table}
\subsection{Formation energies}\label{CeO2pX:ssc_ResultsNV_Ef}
\begin{figure}[!tbp]
\begin{center}
\subfigure{
        \includegraphics[width=8.0cm,keepaspectratio=true]{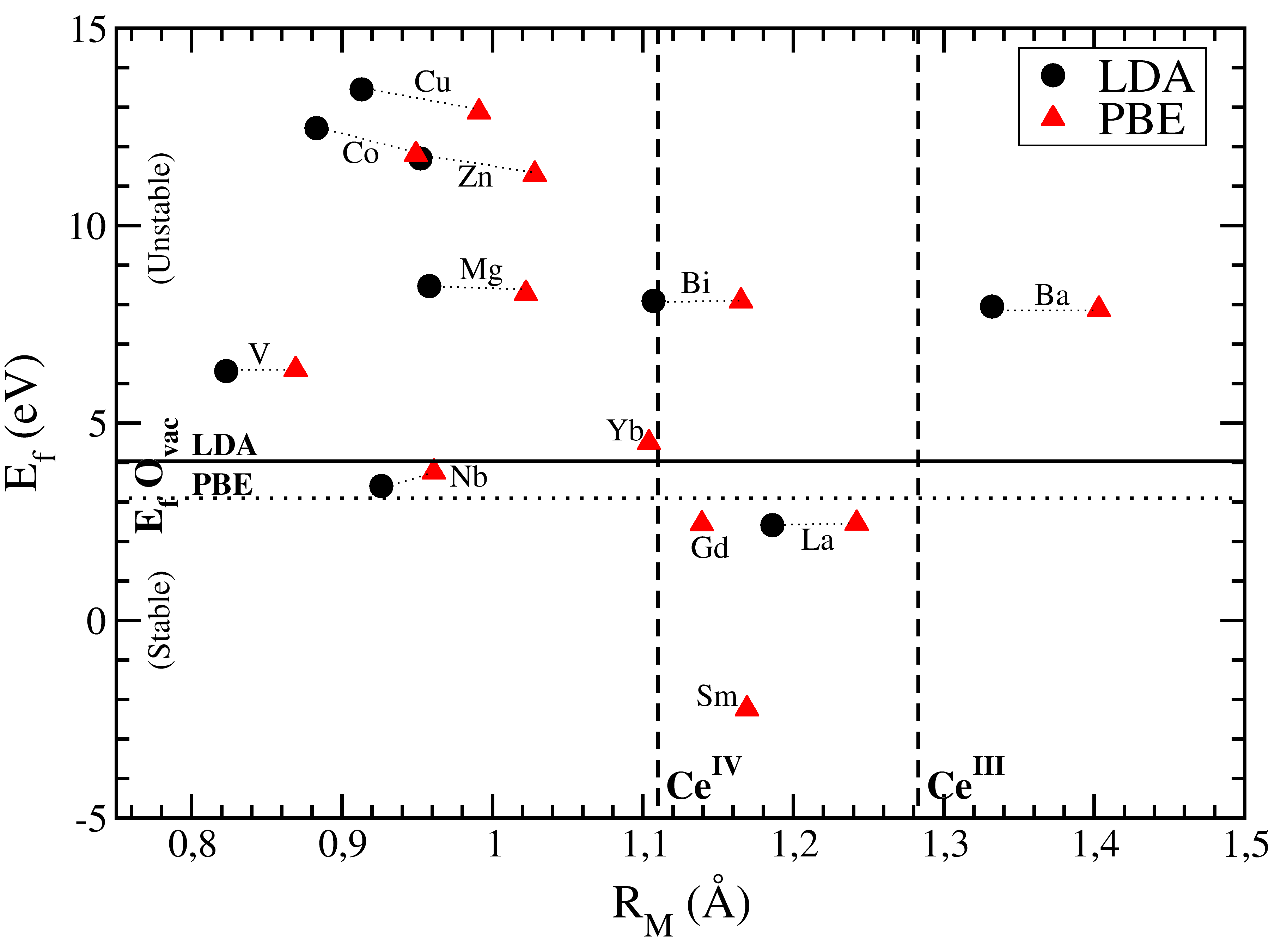}
    \label{fig:subfig:Rm_vs_Ef}
    }
\subfigure{
        \includegraphics[width=8.0cm,keepaspectratio=true]{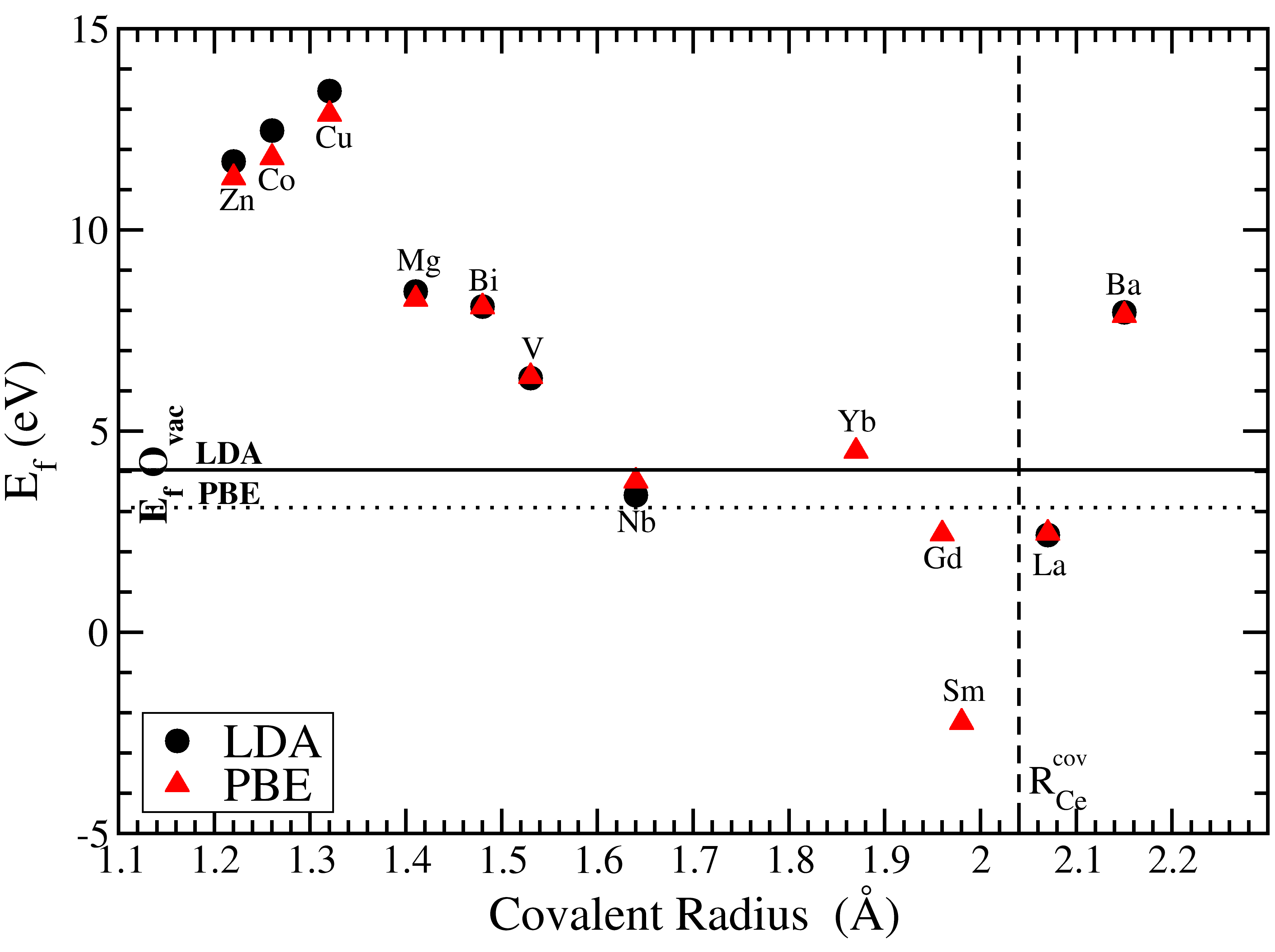}
    \label{fig:subfig:RCov_vs_Ef}
    }
\end{center}
  \caption[Correlation between the formation energy and covalent and calculated crystal radii]{The calculated formation energy E$_f$, for \ce{Ce_{1-$x$}M_{$x$}O2} with $x=0.03125$, as function of the calculated atomic radius R$_M$ (top) and the covalent radius (bottom).\cite{Cordero:DT2008} Top: Vertical dashed lines indicate the Shannon crystal radii for $8-$coordinate \ce{Ce^{III}} and \ce{Ce^{IV}}.\cite{Shannon:table, Shannon:ACSA1976} Bottom: Vertical dashed line indicates the covalent radius for Ce R$^{\mathrm{cov}}_{\mathrm{Ce}}$.\cite{Cordero:DT2008} Top+Bottom: The O vacancy formation energy at a vacancy concentration of $1.5$\% is indicated with a solid (LDA) or dotted (PBE) line.  }\label{fig:R_vs_Ef}
\end{figure}

\indent The stability of the different doped systems is investigated through the comparison of the defect formation energy E$_f$ defined as:
\begin{equation}
E_f = E_{Ce_{1-x}M_{x}O_2} - E_{CeO_2} + N_{df}(E_{Ce} - E_M),
\end{equation}
with $E_{Ce_{1-x}M_{x}O_2}$ the total energy of the doped system, $E_{CeO_2}$ the total energy of a \ce{CeO2} supercell of equal size, $N_{df}$ the number of dopant atoms, and $E_{Ce}$ and $E_M$ the bulk energy per atom of $\alpha$-Ce and the bulk phase of the dopant M. Positive values indicate the amount of energy required to substitute a single Ce atom by a dopant.\\
\indent Defect formation energies given in Table~\ref{table:MetalXsubst_energies} show the same qualitative behavior for the LDA and PBE calculations. Furthermore, as was observed for group IV dopants, formation energies show only limited dependence on the dopant concentration indicating no solubility limits are being crossed within the investigated range.\cite{VanpouckeDanny:2012dGroupIVdopants} The results in Table~\ref{table:MetalXsubst_energies} also show that only Sm doping is stable in an absolute sense with regard to segregation into \ce{CeO2} and bulk \ce{Sm}.\cite{fn:DFTU_Sm} The  positive defect formation energy for the other dopant elements indicates a threshold exists for the formation of these compounds. All the dopants presented in this work have been used in experiments, and of several a \ce{Ce_{1-$x$}M_{$x$}O_{2-$y$}} phase is experimentally observed.\cite{BrisseF:CJChem1967, McBrideJR:JApplPhys1994, TrovarelliA:CatalRev1996, KundakovicLj:ApplCatalA1998, KundakovicLj:JCat1998, MogensenM:SolStateI2000, BeraP:ChemMater2002, RyanKM:JPCondMat2003, BaeJongSung:JoECS2004, KnothKerstin:PhysC2005, deBiasi:JSolStateChem2005, WangX:JPhysChemB2005, WangX:JPhysChemB2006, ManzoliMaela:CT2008, LiB:JPowerSources2008, deBiasi:JAlloysCompd2008, SheYusheng:IJHE2009, VandeVeldeNigel:EurJInorChem2010, LiB:IJHE2010, AiniradA:JAlloysCompd2011, AnwarMS:JAlloysCompd2011, VyshnaviN:2012JMaterChem, TiwariA:ApplPhysLett2006,VodungboB:ApplyPhysLett2007, FernandesV:PhysRevB2007, SongYQ:JApplPhys2007, WenQY:JPhysCondensMatter2007, SinghalRK:JPhysDApplPhys2011, SacanellJ:ApplPhysLett2012, XuD:SolStateIonics2011, HorlaitD:2011InorChem, YaoHC:JPowerSources2012} However, in contrast to the above calculations, experiments are not performed at zero atmosphere and zero Kelvin, and often involve one or more steps which introduce additional energy into the system, providing a means to overcome energy barriers. In addition, the experimental compounds also contain charge compensating vacancies, which are not included in the systems presented in this section. In Sec.~\ref{CeO2pX:sc_IncludeVac}, we will show that the inclusion of such vacancies has only limited influence on the formation energies, allowing the presented defect formation energies to be used as initial indicators of the system stability.\\
\indent Since the formation energies presented here spread over quite a wide range it is obvious that not all dopants will form a compound system equally easily. In consequence, a reference is needed to indicate which are more likely to form a doped bulk phase and which dopants are more likely to segregate (to the surface in case of for example catalyst nanocrystals). It is well-known for \ce{CeO2} to spontaneously form oxygen vacancies, so the oxygen vacancy formation energy of pure \ce{CeO2} can be used as a reference for the likelihood of forming a \ce{Ce_{1-$x$}M_{$x$}O_{2}} bulk-phase.\cite{TrovarelliA:CatalRev1996, MogensenM:SolStateI2000} Table~\ref{table:MetalXsubst_vac_Ef} shows the calculated oxygen vacancy formation energy for \ce{CeO_{1.96875}} to be $4.035$ and $3.097$ eV for LDA and PBE, respectively, going up to $5.006$ and $4.145$ eV in \ce{CeO_{1.75}}. From this we conclude that Nb and the lanthanides presented in this work are likely to form fluorite based bulk-phases of \ce{Ce_{1-$x$}M_{$x$}O2}, while the other dopants are expected to segregate either into internal domains or to the surface of the grains. However, combining this knowledge, with the results of the calculated dopant radii, also points toward another option: a high defect formation energy and a crystal radius indicative of a preference for lower coordination may indicate that local reconstructions around the dopant are present in experiment. Such reconstructions would lead to a better suited chemical environment, with better matched coordination, and should give rise to lower defect formation energies.\\
\indent Of all dopants presented in this work, Cu shows the highest formation energy, making it the most likely candidate for phase segregation and/or reconstruction. The existence of such reconstruction is shown in the work of Wang \textit{et al.}\cite{WangX:JPhysChemB2005} and Lu \textit{et al.}\cite{LuZ:ChemPhysLett2011} where a symmetry breaking reconstruction for the Cu dopant was found and investigated. But even when this reconstructed structure is taken into account, Cu doped \ce{CeO2} remains one of the most unstable systems. The tetragonal reconstruction is only $1.223$ eV more stable than the cubic fluorite structure, resulting in a defect formation energy of about $11.7$ eV in PBE calculations. In the literature several experimental groups have investigated \ce{CuO} doped/modified \ce{CeO2} showing a general trend of phase segregation for medium to high Cu content.\cite{KundakovicLj:ApplCatalA1998, KundakovicLj:JCat1998, BeraP:ChemMater2002, LinR:ApplCatal2003, WangX:JPhysChemB2005, WangX:JPhysChemB2006, deBiasi:JAlloysCompd2008, SheYusheng:IJHE2009} Kundakovic and Flytzani-Stephanopoulos investigated the reduction characteristics of \ce{CuO} dispersed on \ce{Ce_{1-$x$}La_{$x$}O2} catalyst supports.\cite{KundakovicLj:ApplCatalA1998, KundakovicLj:JCat1998} They found that for low Cu content, copper is present as small clusters or even isolated ions. For higher concentrations, also \ce{CuO} particles are observed.\cite{KundakovicLj:ApplCatalA1998} Similar observations have been reported by Lin \textit{et al.} and also de Biassi and Grillo present evidence of Cu clustering.\cite{LinR:ApplCatal2003, deBiasi:JAlloysCompd2008} In addition, Kundakovic and Flytzani-Stephanopoulos also present the observation of bulk doped \ce{Ce_{0.99}Cu_{0.01}O_{2-$y$}} for calcination temperatures below $500^{\circ}$C, and state that for higher calcination temperatures the Cu ions segregate to the surface to form clusters. This supports the instability of Cu doped \ce{CeO2} predicted by our calculated formation energies.\\
\indent In contrast, Bera \textit{et al.} do not observe any \ce{CuO} related lines in their X-ray diffraction spectra for $3-5$\% Cu doping, nor do they observe \ce{CuO} particles in their TEM measurements. As a result they conclude Cu ions to be present in the \ce{CeO2} crystal matrix. However, they also note that there are $4$ to $6$ times as many Cu ions located on the surface of the \ce{CeO2} particles.\cite{BeraP:ChemMater2002} Combined with the results for low calcination temperatures of Kundakovic and Flytzani-Stephanopoulos this would appear to indicate that a significant kinetic barrier is present for the Cu ions, limiting the mobility of the Cu ions somewhat, which in turn also limits their ability for clustering and/or segregation to the surface after their dispersal in the \ce{CeO2} bulk during for example high temperature treatment.\\
\indent In contrast, Co bulk-doping, which is shown in Table~\ref{table:MetalXsubst_energies} to be almost as unfavorable as Cu doping, is widely used in experimental studies in the context of dilute semiconductors. In many of these experiments, samples which are often thin films, are prepared via pulsed laser deposition.\cite{SongYQ:JApplPhys2007, VodungboB:ApplyPhysLett2007, TiwariA:ApplPhysLett2006} Observation of \ce{Ce_{1-$x$}Co_{$x$}O_{2-$y$}} in these samples may be an indication that quite high kinetic barriers are present, effectively pinning the Co ions in place despite the unfavorable energetics. Alternatively, Co may segregate into very small \ce{Co}/\ce{CoO} clusters, which could at higher Co concentrations give rise to the \ce{Co3O4} impurities observed by Sacanell et al.\cite{SacanellJ:ApplPhysLett2012} This would also be in line with the calculated preference of oxygen vacancies near Co ions.\cite{MurgidaGE:SolidStateComm2012} Yao \textit{et al.} investigated the codoping of \ce{CeO2} with Co and Sm.\cite{YaoHC:JPowerSources2012} They observe no evident secondary \ce{CoO} and \ce{Co3O4} phases. In combination with the observation of Vegard law behavior as function of the Co content, they conclude the Co ions to be incorporated into the ceria lattice to form a solid solution. This seems to indicate that the codoping with Sm in this case stabilizes the Co dopants somewhat, which is not unreasonable based on the Sm defect formation energy given in Table~\ref{table:MetalXsubst_energies}. In addition, Yao \textit{et al.} also observe the grain boundary conductivity to show a maximum at $5$\% of Co doping.\cite{YaoHC:JPowerSources2012} They link this to the segregation of Co to the grain boundary, showing that the Sm dopants can only stabilize a limited amount of Co.\\
\indent Another interesting dopant to have a closer look at is \ce{Ba}. Of the dopants investigated, it shows the largest decrease in defect formation energy for increasing concentration. Combined with the relatively large defect formation energy this could be an indication that the \ce{BaCeO3} interface observed between superconducting \ce{YBa2Cu3O_{7-$\delta$}} thin films and \ce{CeO2} buffer layers is rather due to Ce moving into \ce{BaO} layers than Ba moving into the \ce{CeO2} buffer layer.\cite{TakahashiY:PhysC2004, VandeVeldeNigel:EurJInorChem2012} On the other hand, if it are the Ba atoms that diffuse into the \ce{CeO2} buffer layer, then doping of the \ce{CeO2} buffer layer with dopants that have a lower defect formation energy may prevent the Ba diffusion by blocking possible sites. However, before any conclusive statement can be made further theoretical work is be needed; for example a comparative study of Ce doping of bulk \ce{BaO} or \ce{BaO} layers in \ce{YBa2Cu3O_{7-$\delta$}} and Ba doping in \ce{CeO2} or doped \ce{CeO2}. This is, however, beyond the scope of the current work.\\
\indent In Fig.~\ref{fig:R_vs_Ef} the obtained  defect formation energies are compared to the calculated atomic crystal radius R$_M$ (Fig.~\ref{fig:subfig:Rm_vs_Ef}) and the covalent radius (Fig.~\ref{fig:subfig:RCov_vs_Ef}).\cite{Cordero:DT2008}  Figure~\ref{fig:subfig:Rm_vs_Ef} shows the most stable dopants to have a crystal radius between that of $8-$coordinate \ce{Ce^{IV}} and \ce{Ce^{III}}, while Fig.~\ref{fig:subfig:RCov_vs_Ef} shows high stability for elements with a covalent radius close to that of Ce. In both cases, the Nb dopant appears as an exception, showing a reasonably beneficial defect formation energy, while presenting a significantly lower atomic radius than the other more stable dopants. The Nb covalent radius, however, is nicely in the range of those of the group IVb elements (Ti: $1.60$\AA, Zr and Hf: $1.75$\AA) which were shown to provide stable dopants.\cite{Cordero:DT2008, VanpouckeDanny:2012dGroupIVdopants} The same is true for the calculated atomic radii R$_M$.\cite{VanpouckeDannyEP:2012aApplSurfSci} The main difference between Nb and the other elements presented here is the fact that it acts as a tetravalent dopant in a \ce{Ce_{1-$x$}Nb_{$x$}O2} system. This shows that the relation between dopant stability and radius is more complex, and that the oxidation state (in the compound) plays an important role. As such, a higher oxidation state results in smaller radii for stable dopants. In addition, it is also apparent from these figures that the order of the atomic radii differs significantly depending of the definition used. Consequently, simple stability rules based on ratios of atomic radii, for example used in the study of the fluorite-pyrochlore transitions, should be treated with considerable caution since they appear to be ill-defined.\cite{MinerviniL:JAmCeramSoc2000, YamamuraH:JCSJ2004}\\
\indent Based on the results presented in this work, it is also possible to make some predictions about other dopant elements. Let us assume that the trend observed for the defect formation energies of group IVa and IVb dopants also hold for other groups.\cite{VanpouckeDanny:2012dGroupIVdopants} Then, from the defect formation energies for Mg and Ba, we can conclude that all elements of the group IIa will segregate when used as a dopant in \ce{CeO2}. From the values calculated for V and Nb on the other hand, the value for Ta is expected to be below the oxygen vacancy formation energy, indicating Ta to be a good dopant candidate for presenting a meta-stable bulk phase. This is supported by the experimental work of Zhao and Gorte, who studied the influence of \ce{Ta2O5} doping of \ce{CeO2} on its catalytic activity for $n$-butane.

\begin{table}[!bt]
\caption[Bulk moduli and thermal expansion coefficients for aliovalent dopants]{Calculated BM $B_0$ for \ce{CeO2} at a dopant concentration of $25$\%, for LDA and PBE calculations. The linear TEC $\alpha$ at the same dopant concentration and a temperature of $500$ K.\cite{fn:missLDA} A best guess for the oxidation state (ox.) of the dopants is given.\cite{fn:ValenceDisclaimer}}\label{table:MetalXsubst_BM_TEC}
\begin{ruledtabular}
\begin{tabular}{l|c|cccc}
& ox. & \multicolumn{2}{c}{$B_0$ (Mbar)} &  \multicolumn{2}{c}{$\alpha$ ($10^{-6}$ K$^{-1}$)} \\
& & LDA & PBE & LDA & PBE \\
\hline
\ce{CeO2} &  & $2.017$ & $1.715$ & $11.218$ & $12.955$ \\
\hline
Mg & II  & $1.644$ & $1.389$ & $14.693$ & $16.867$ \\
V  & IV  & $2.132$ & $1.796$ & $11.510$ & $13.601$ \\
Co & II  & $1.867$ & $1.542$ & $13.779$ & $16.567$ \\
Cu & I   & $1.704$ & $1.374$ & $16.186$ & $19.902$ \\
Zn & II  & $1.712$ & $1.410$ & $14.968$ & $17.656$ \\
Nb & IV  & $2.187$ & $1.871$ & $10.621$ & $12.226$ \\
Ba & II  & $1.580$ & $1.321$ & $13.544$ & $15.608$ \\
La & III & $1.835$ & $1.556$ & $11.809$ & $13.618$ \\
Sm & III & $-$     & $1.595$ & $-$      & $13.678$ \\
Gd & III & $-$     & $1.588$ & $-$      & $13.744$ \\
Yb & III & $-$     & $1.534$ & $-$      & $15.229$ \\
Bi & III & $1.874$ & $1.575$ & $12.631$ & $14.836$ \\
\end{tabular}
\end{ruledtabular}
\end{table}
\subsection{Bulk modulus and Thermal expansion coefficients}\label{CeO2pX:ssc_ResultsNV_BM_TEC}
\begin{figure}[!tb]
\begin{center}
    \includegraphics[width=8cm,keepaspectratio=true]{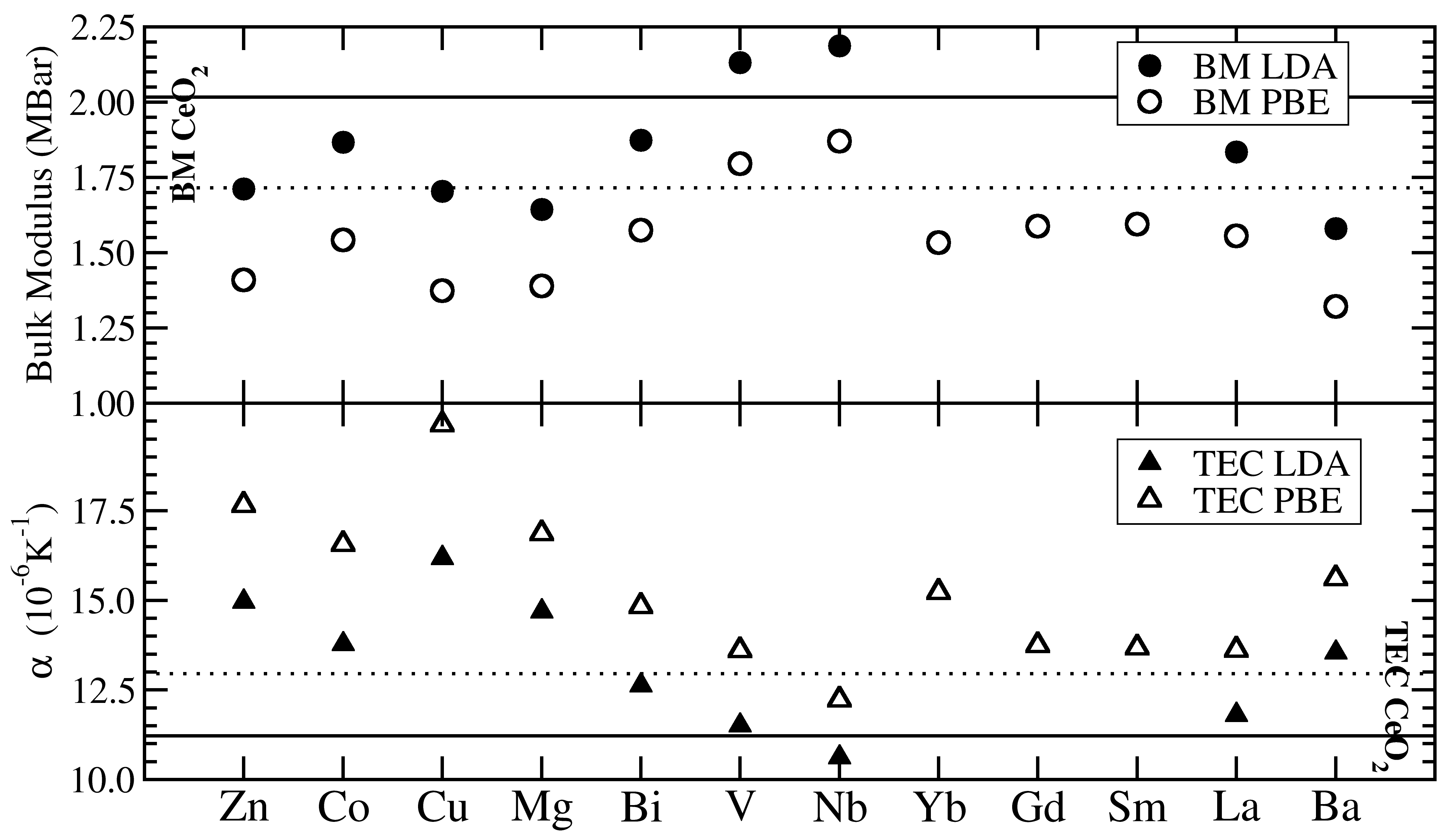}\\
\end{center}
  \caption[Comparison of bulk moduli and thermal expansion coefficients for \ce{Ce_{0.75}M_{0.25}O2}]{Calculated BM and linear TEC $\alpha$ at $500$ K for \ce{Ce_{0.75}M_{0.25}O2}. Calculated values (LDA: solid line, PBE: dotted line) for pure \ce{CeO2} are given for reference. The elements are sorted with regard to increasing Covalent radius.\cite{Cordero:DT2008}}\label{fig:BM_vs_TEC}
\end{figure}

\begin{figure}[tb]
\begin{center}
    \includegraphics[width=8cm,keepaspectratio=true]{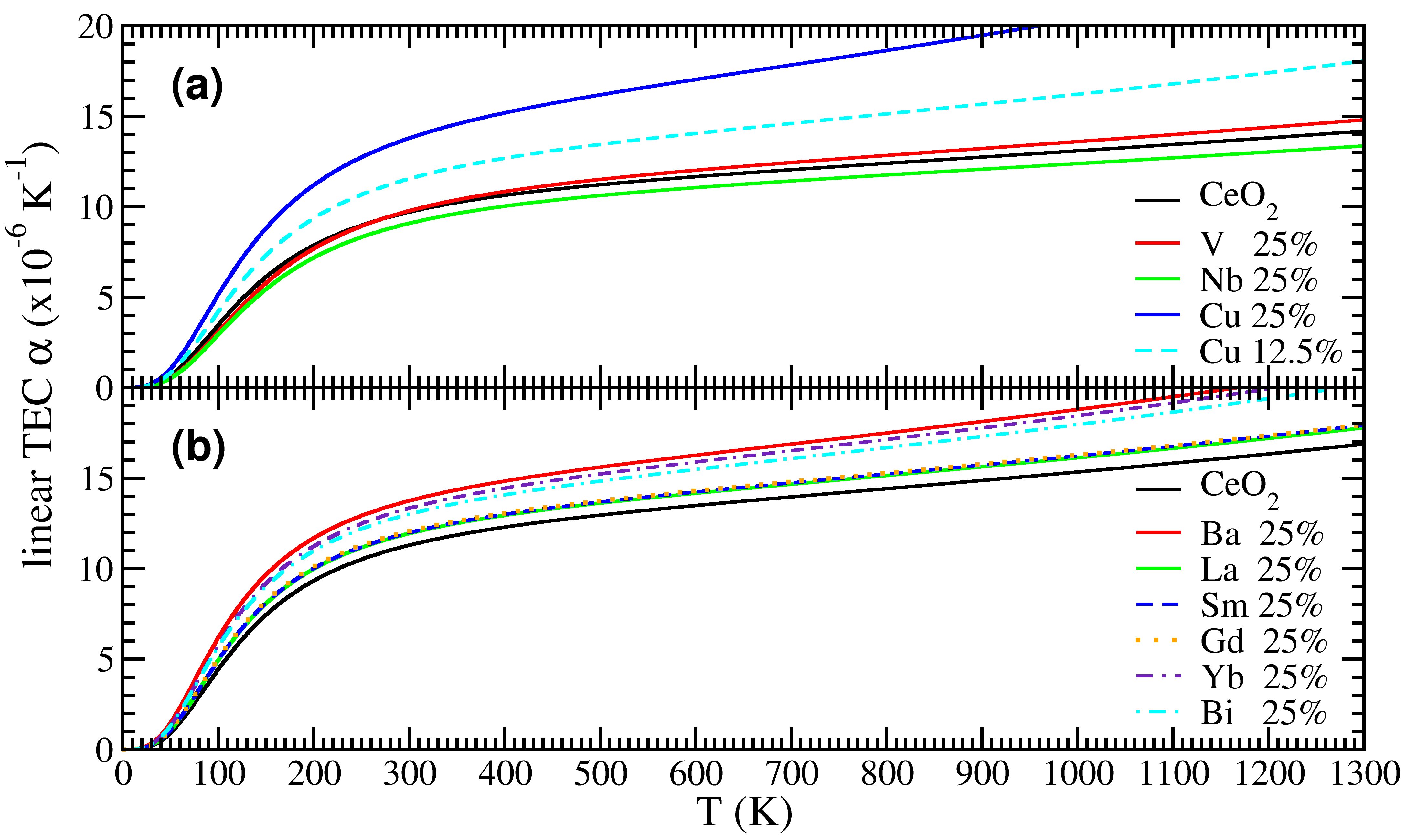}\\
\end{center}
  \caption[Calculated linear thermal expansion coefficient $\alpha$ for different dopants]{Calculated linear TEC $\alpha$ for different dopants based on (a) LDA and (b) PBE total energies and volumes. The calculated TEC of \ce{CeO2} (black solid curve) is given as reference.}\label{fig:TECplotNV}
\end{figure}

\indent The modification of the elastic properties of \ce{CeO2} due to aliovalent doping is investigated through the BM and linear TEC $\alpha$. To reduce the computational cost, BM and TEC are only calculated for dopant concentrations of $25$\%. Table~\ref{table:MetalXsubst_BM_TEC} shows the BM and the linear TEC at $500$ K. The BM and TEC for pure \ce{CeO2} are given as reference. These show the LDA based value for the TEC to be in excellent agreement with the experimental value ( $(11.0\pm 0.5)\times 10^{-6}$\ K$^{-1}$ at room temperature (RT), and $(11.5\pm 0.5)\times 10^{-6}$\ K$^{-1}$ at $500\ ^{\circ}$C ), while the PBE value is clearly an over-estimation.\cite{MogensenM:SolStateI2000} With regard to the BM it is again the LDA value which shows best agreement with experiment where values in the range of $2.04$--$2.36$MBar have been measured.\cite{GerwardL:JAlloysCompd2005, DuclosSJ:RhysRevB1988, NakajimaA:PhysRevB1994} The PBE value shows a significant underestimation, in line with the overestimation of the TEC. Of all dopants investigated here, only V and Nb give rise to an increase in the BM, all other dopants reduce the BM to varying degree. Comparing the BM for (the tetravalent) V and Nb dopants to those found for group IVb dopants shows them to present similar values.\cite{VanpouckeDannyEP:2012aApplSurfSci, VanpouckeDanny:2012dGroupIVdopants} For Cu the BM was also calculated for \ce{Ce_{0.875}Cu_{0.125}O2}, and found to be $1.867$ and $1.553$ Mbar for LDA and PBE, respectively. This is within $0.01$ Mbar of the average of the BM for pure \ce{CeO2} and \ce{Ce_{0.75}Cu_{0.25}O2}, showing that a linear relation between the BM and dopant concentration is a reasonable assumption for \ce{Ce_{1-$x$}M_{$x$}O2} systems.\\
\indent With the exception of Nb, all investigated dopants result in an increase of the TEC. The data in Table~\ref{table:MetalXsubst_BM_TEC} reveal that low dopant valence leads to a large increase in the TEC and high valence leads to a small increase and even decrease of the TEC.\\
\indent Comparison of the BM and the TEC in Fig.~\ref{fig:BM_vs_TEC} shows clearly opposite trends of the BM and TEC, as was also observed for group IV elements, again showing the expected inverse correlation between the BM and the TEC.\cite{VanpouckeDanny:2012dGroupIVdopants, TsuruY:JCeramJpn2010} Only vanadium shows a slightly different behavior with both the TEC and BM being larger than the \ce{CeO2} values. Close investigation of the vanadium TEC in Fig.~\ref{fig:TECplotNV}a shows that the vanadium curve crosses the TEC curve for pure \ce{CeO2} at around $250$ K, so below this temperature the inverse behavior of the TEC and BM is restored.\\
\indent Figure~\ref{fig:TECplotNV}a also shows the TEC for two different Cu dopant concentrations. In the range of roughly RT up to at least $1000$ K a nearly linear influence of the dopant concentration on the TEC change is observed, indicating that for aliovalent dopants in highly oxidizing atmosphere the TEC may also be linearly interpolated. This linear behavior supports the inherent assumption underlying the experimental practice of codoping in several ceria based applications.\cite{FaggDP:JSolStateChem2006, AiniradA:JAlloysCompd2011}\\
\indent Figure~\ref{fig:TECplotNV}b shows the TEC of the lanthanides La, Sm, and Gd to coincide nicely, while the Yb curve shows much higher values. This difference in behavior is most likely linked to the filled $4f$ shell of Yb (which is only partially filled for Sm and Gd). Further investigation of lanthanide dopants is required to have the full picture of the mechanism at work. Similar as was found for group IV dopants, this behavior shows the importance of filled shells near the Fermi-level.\\
\indent With regard to the comparison of calculated and experimentally obtained lattice parameters for \ce{CeO2}, several authors have noted that one should be very careful, since the former are generally calculated at zero Kelvin, while the latter are measured at RT. These authors suggest to linearly extrapolate the calculated lattice parameter making use of the `\textit{linear TEC}'. In this setup the coefficient is assumed to be a constant, and often taken from experiment. As is shown in Fig.~\ref{fig:TECplotNV}, the linear TEC shows quite a non-linear behavior at low temperature.\cite{fn:linTEC} Taking this behavior into account one can obtain a more accurate value of the lattice parameter at RT. Zero Kelvin and RT values of the lattice parameter of doped \ce{CeO2} are shown in Table~\ref{table:MetalX_Rm_Vegards_aRT}. The thermal contribution to the lattice parameter at RT is fairly limited and is of the order of $0.02$--$0.04$ \AA, for dopant concentrations of $25$\%. Since this can be comparable to the lattice parameter change due to the introduction of a dopant, this can result in different doped systems to have the same lattice parameter at elevated temperatures (e.g. Sm and Bi doped ($25$\%) \ce{CeO2} at about $1065$ K, and pure and Yb doped ($25$\%) \ce{CeO2} at about $1024$ K). As a result, codoped systems or interfaces between layers of differently doped \ce{CeO2} may experience reduced strain at elevated temperatures. The opposite is to be expected as well, and increased segregation or interface strain at elevated temperatures could be a consequence. This latter aspect is of importance when perfect interfaces are required, and should be considered when crack formation in thin films is an issue.\cite{ParanthamanM:1997PhysC, OhSanghyun:PhysC1998, TakahashiY:PhysC2004, VandeVeldeNigel:EurJInorChem2010}
\section{Inclusion of Vacancies}\label{CeO2pX:sc_IncludeVac}
\indent Since the configuration of dopants and oxygen vacancies for the systems studied is essentially unknown, different configurations need to be investigated. However, since it is neither our goal nor our intent to find the exact ground state configuration of these systems, but rather to investigate the influence of vacancies, we will restrict ourselves to a subset of dopants and a small set of configurations for the different dopants. A full study of the configurational space is beyond the scope of this study. The subset of dopants consists of \ce{Cu^{I}}, \ce{Zn^{II}}, and \ce{Gd^{III}}. In addition, vacancies in pure \ce{CeO2} are added as reference.\\
\indent The different vacancy geometries are described in Sec.~\ref{CeO2pX:sc_theormeth} and the notation `NV' is used to indicate the `No Vacancy' reference systems, \textit{i.e.} \ce{Ce_{1-$x$}M_{$x$}O2} with M$=$\ce{Cu}, \ce{Zn}, or \ce{Gd}. All systems presented, contain $1$ oxygen vacancy per dopant atom, resulting in charge under-compensation (Cu), compensation (Zn), and over-compensation (Gd).\cite{fn:Alio:CompChargeVacDef} For these systems a vacancy formation energy E$_{vac}$ is calculated as:
\begin{equation}\label{eq:Evac_def}
\mathrm{E}_{vac} = E_{Ce_{1-x}M_{x}O_{2-y}} +\frac{N_{vac}}{2}E_{O_2}- E_{Ce_{1-x}M_{x}O_2},
\end{equation}
with $N_{vac}$ the number of oxygen vacancies,\cite{fn:VacNrDef} $E_{O_2}$ the total energy of an oxygen molecule, and $E_{Ce_{1-x}M_{x}O_{2-y}}$ and $E_{Ce_{1-x}M_{x}O_2}$ the total energies of the system with and without vacancies, respectively.
\subsection[Oxygen and cerium vacancies in \texorpdfstring{\ce{CeO2}}{CeO{\texttwoinferior}}]{Oxygen and cerium vacancies in \ce{CeO2}.}\label{CeO2pX:sscandCe_vac}
\begin{figure}[tb]
\begin{center}
    \includegraphics[width=8cm,keepaspectratio=true]{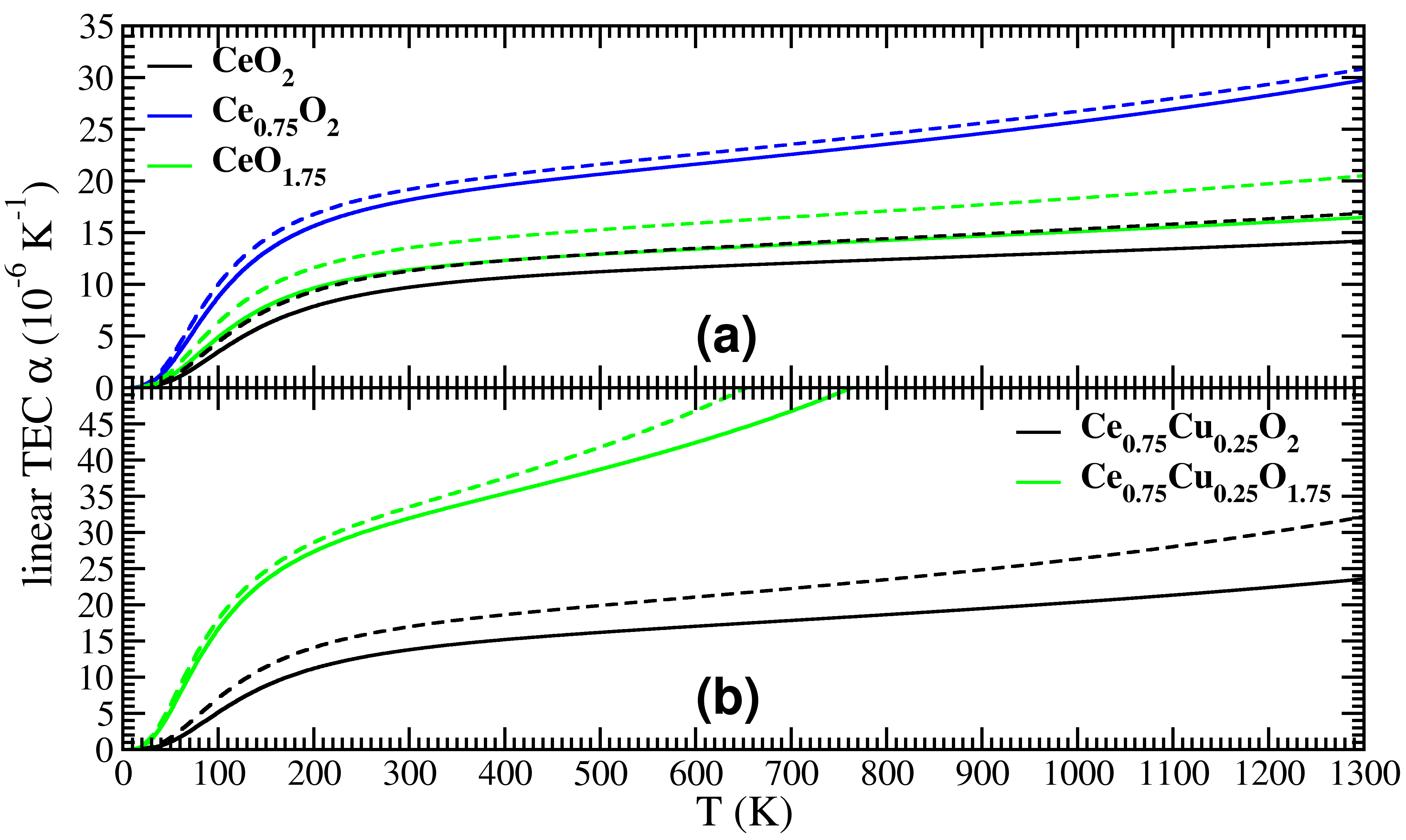}\\
\end{center}
  \caption[Thermal expansion coefficients for doped systems containing oxygen vacancies]{Calculated linear TEC $\alpha$ for different systems containing vacancies. (a) Comparison of the influence of oxygen and cerium vacancies, (b) Cu doping with and without oxygen vacancies. LDA results are shown as solid lines and PBE as dashed lines.
  }\label{fig:TECplotVac}
\end{figure}

\indent Before investigating the combined influence of dopants and vacancies, the influence of oxygen and cerium vacancies on pure \ce{CeO2} is briefly discussed. Table~\ref{table:MetalXsubst_vac_Ef} shows the vacancy formation energy of both Ce and O vacancies. From this, it is clear that Ce vacancies are highly unfavorable, in agreement with experimental observations.\cite{TrovarelliA:CatalRev1996, MogensenM:SolStateI2000} In addition, the relatively small change of the lattice parameter appears to be strongly functional dependent.\\
\indent The vacancy formation energy of the oxygen vacancies on the other hand shows a significant concentration dependence (in contrast to the dopant calculations of the previous section). In addition, the calculated lattice expansion is clearly non-linear, with a similar trend for LDA and PBE calculations. The expansion of the lattice parameter due to the presence of oxygen vacancies is experimentally known, and theoretically understood as a consequence of the transition from \ce{Ce^{IV}} to \ce{Ce^{III}} of two Ce atoms neighboring the oxygen vacancy. Since the atomic crystal radius of \ce{Ce^{III}} is significantly larger than \ce{Ce^{IV}} ($1.283$\AA\ instead of $1.11$\AA) the lattice will expand.\cite{TrovarelliA:CatalRev1996, MogensenM:SolStateI2000, Shannon:ACSA1976, Shannon:table} The non-linearity shown here, indicates that for aliovalent dopants charge compensating vacancies may give rise to non-Vegard law behavior, due to \ce{Ce^{IV} -> Ce^{III}} transitions.\\
\indent Interesting to note is the large impact of the vacancies on the BM and TEC of \ce{CeO2}. Figure~\ref{fig:TECplotVac}a shows a dramatic increase in the linear TEC due to the presence of vacancies. It is clear that the inverse relation between the BM and the TEC is retained for vacancies.
\begin{table*}[!bt]
\caption[Properties of vacancies in non-doped \ce{CeO2}]{Properties of vacancies in non-doped \ce{CeO2}: vacancy formation energy $E_{vac}$ as given in Eq.~\eqref{eq:Evac_def}, lattice expansion $\Delta a_0$, bulk modulus $B_0$ and linear thermal expansion coefficient $\alpha$. Vacancy concentrations are indicated and the linear thermal expansion coefficient value $\alpha$ is given for a temperature of $500$ K.}\label{table:MetalXsubst_vac_Ef}
\begin{ruledtabular}
\begin{tabular}{ll|cccc|cccc|cc}
 & & \multicolumn{4}{c|}{E$_{vac}$} & \multicolumn{4}{c|}{$\Delta a_0$} & $B_0$ & $\alpha$ \\
 & & \multicolumn{4}{c|}{(eV)} & \multicolumn{4}{c|}{(\%)} & (Mbar) & ($10^{-6}$ K$^{-1}$) \\
\hline
 \multicolumn{2}{l|}{Vac. conc. (\%)} &  $12.5$\% & $6.25$\% & $1.852$\% & $1.563$\% & $12.5$\% & $6.25$\% & $1.852$\% & $1.563$\% & $12.5$\% & $12.5$\%\\
\hline
\multirow{2}{*}{O Vac.}& LDA & $5.006$ & $4.440$ & $4.054$ & $4.035$ & $0.775$ & $0.510$ & $0.176$ & $0.141$ & $1.568$ & $12.912$ \\
                       & PBE& $4.145$ & $3.476$ & $3.075$ & $3.097$ & $0.908$& $0.606$ & $0.193$ & $0.165$ & $1.320$ & $15.287$ \\
\hline
 \multicolumn{2}{l|}{Vac. conc. (\%)} &  $25$\% & $12.5$\% & $3.704$\% & $3.125$\% & $25$\% & $12.5$\% & $3.704$\% & $3.125$\% & $25$\% & $25$\%\\
\hline
\multirow{2}{*}{Ce Vac.}& LDA & $17.549$ & $17.779$ & $17.857$ & $17.829$ & $-0.270$ & $-0.032$ & $-0.016$ & $-0.028$ & $1.023$ & $20.650$ \\
& PBE& $16.255$ & $16.543$ & $16.611$ & $16.592$ & $0.560$ & $0.266$ & $0.063$ & $0.061$ & $0.858$ & $21.609$ \\
\end{tabular}
\end{ruledtabular}
\end{table*}

\begin{table*}[!bt]
\caption[Properties for Cu, Zn and Gd doped \ce{CeO2} containing oxygen vacancies]{Calculated vacancy formation energy (E$_{vac}$) as given by Eq.~\eqref{eq:Evac_def}, bulk modulus ($B_0$), and change in volume ($\Delta V$) and lattice expansion ($\Delta a_0$) for Cu, Zn, and Gd doped \ce{CeO2} including a single vacancy per supercell. $\Delta V$ and $\Delta a_0$ are taken with regard to pure \ce{CeO2}. All calculations are performed using PBE functionals. Vacancy concentrations are $12.5$ (c111) and $6.25$\% (p222). The different configurations are shown in Fig.~\ref{fig:c111p222_geom}. NV indicates the reference systems without vacancies. }\label{table:MetalXsubst_vac_Ef_BM_dV_da}
\begin{ruledtabular}
\begin{tabular}{l|ccc|ccc|ccc|ccc}
& \multicolumn{3}{c|}{E$_{vac}$ (eV)} &  \multicolumn{3}{c|}{$B_0$ (Mbar)} &  \multicolumn{3}{c|}{$\Delta V$ (\%)} &  \multicolumn{3}{c}{$\Delta a_0$ (\%)}\\
&  Cu & Zn & Gd &  Cu & Zn & Gd &  Cu & Zn & Gd &  Cu & Zn & Gd \\
\hline
c$111$ NV               & $12.922^{a}$ & $11.057^{a}$ & $2.396^{a}$ & $1.37$ & $1.41$ & $1.59$ & $-4.124$ & $-3.189$ & $0.274$ & $-1.394$ & $-1.074$ & $0.091$ \\
c$111$ V$_{\mathrm{B}}$ & $-0.800$ & $-0.882$ & $1.929$ & $0.87$ & $0.38$ & $0.86$ & $-0.568$ & $-1.529$ & $1.563$ & $-0.190$ & $-0.512$ & $0.518$ \\
\hline
p$222$ NV               & $12.878^{a}$ & $11.249^{a}$ & $2.445^{a}$ & $1.55$ & -- & -- & $-2.145$ & $-1.529$ & $0.158$ & $-0.733$ & $-0.526$ & $0.039$ \\
p$222$ V$_{\mathrm{A}}$ & $-0.048$ & $-0.604$ & $1.395$ & $1.01$ & $1.03$ & $1.00$ & $\phantom{-}0.619$ & $\phantom{-}0.552$ & $1.371$ & $\phantom{-}0.193$ & $\phantom{-}0.170$ & $0.442$ \\
p$222$ V$_{\mathrm{B}}$ & $-0.422$ & $-1.200$ & $1.364$ & $1.04$ & $1.09$ & $1.31$ & $-1.173$ & $-1.416$ & $1.116$ & $-0.406$ & $-0.488$ & $0.357$ \\
\end{tabular}
\end{ruledtabular}
\begin{flushleft}
$^{a}$ For the systems without vacancies, the formation energy E$_{f}$ is repeated. Note that the formation energy of a doped system with oxygen vacancies E$_{f,vac} =$E$_{f} + $E$_{vac}$.\\
\end{flushleft}
\end{table*}

\subsection[Aliovalent dopants + a single vacancy]{Aliovalent dopants Cu, Zn and Gd combined with a single oxygen vacancy.}\label{Alio:ssc_alioMplusVac}
\setcounter{paragraph}{0} 
\paragraph{Vacancy formation energy.\\}
\indent If one assumes the oxidation states of Cu, Zn and Gd as dopants for \ce{CeO2} to be I, II, and III, respectively, then the introduction of a single oxygen vacancy for every dopant ion will result in under-compensation in case of Cu, nominal charge compensation for Zn, and over-compensation in case of Gd.\cite{fn:ValenceDisclaimer} Table~\ref{table:MetalXsubst_vac_Ef_BM_dV_da} shows the vacancy formation energies for these three dopants. For all systems, the absolute value of E$_{vac}$ is of the order of $1$ eV. Because the formation energy of a doped system including oxygen vacancies can be written as
E$_{f,vac} =$E$_{f} + $E$_{vac}$, where E$_{f}$ is the formation energy of the NV system, the introduction of an oxygen vacancy in a Cu or Zn doped system will result in an improved stability. However, since E$_{f,vac}$ is positive this means that the formation of oxygen vacancies will not prevent phase segregation and promote the formation of bulk doped \ce{CeO2}. This would require E$_{vac}$ to be more negative than E$_{f}$ is positive.\\
\indent In contrast, the Gd doped system appears to destabilize due to the introduced oxygen vacancy. This destabilization is merely a consequence of the fact that the vacancy concentration is higher than the nominal concentration required for charge compensation. Table~\ref{table:MetalXsubst_Gd05Vac_Ef} shows the vacancy formation energies for different configurations containing two Gd dopant ions and a single vacancy leading to exact charge compensation. In this situation, the vacancies also have a stabilizing effect on the Gd doped system. Note that the effect of different configurations without vacancies is quite small, all have defect formation energies within a range of $50$ meV. The oxygen vacancy formation energies on the other hand are spread over a wider range, and show a correlation with chemical environment defined as the surrounding cation tetrahedron (\textit{cf.}~Fig.~\ref{fig:c111p222_geom}a). The oxygen vacancy appears to prefer multiple dopant cations in the tetrahedral surrounding in case of Gd doping. Based on the A and B configurations, there appears to be an improvement of the vacancy formation energy of $150$ meV per Gd cation included in the tetrahedron. This shows good agreement with the association energy of $0.13$ eV for the Gd-oxygen vacancy complex.\cite{SteeleBCH:SolStateIon2000} This is in line with earlier atomistic calculations of Catlow and collaborators, and supports the predicted instability of a pyrochlore phase for \ce{Ce2Gd2O7} by Minervini and collaborators.\cite{CatlowCRA:SolidStateIonics1983, ButlerV:SolStateIonics1983, MinerviniL:JAmCeramSoc2000} For La, which is also a trivalent dopant for \ce{CeO2}, an opposite trend was noted for the $50$\% doped system.\cite{VanpouckeDannyEP:2011PhysRevB_LCO}
\begin{table}[!bt]
\caption[Oxygen vacancy formation energies for several \ce{Ce_{0.75}Gd_{0.25}O_{1.875}} configurations]{Oxygen vacancy formation energies for different \ce{Ce_{0.75}Gd_{0.25}O_{1.875}} configurations. NV indicates configurations without oxygen vacancies. The different configurations are shown in Fig.~\ref{fig:c111A1x_Gd_geom}. The number of Gd ions in the tetrahedron surrounding the vacancy is given.}\label{table:MetalXsubst_Gd05Vac_Ef}
\begin{ruledtabular}
\begin{tabular}{l|cccc|c}
& \multicolumn{4}{c|}{E$_{vac}$ (eV)} & $\Delta V (\%)$\\
& NV$^{a}$ & V$_{A}$ & V$_{B}$ & V$_{C}$ & V$_{B}$ \\
\# Gd & na & $0$ & $1$ & $2$ & $1$ \\
\hline
A & $2.380$ & $-0.137$ & $-0.273$ & $-0.440$ & $0.943$ \\
B & $2.365$ & $-0.037$ & --       & $-0.310$ & --\\
C & $2.398$ & --       & $-0.392$ & --       & $0.871$ \\
D & $2.402$ & --       & $-0.402$ & --       & $0.866$ \\
\end{tabular}
\end{ruledtabular}
\begin{flushleft}
$^{a}$ For the systems without vacancies, the formation energy E$_{f}$ is presented. Note that the formation energy of a doped system with oxygen vacancies E$_{f,vac} =$E$_{f} + $E$_{vac}$.\\
\end{flushleft}
\end{table}
\indent Also for Cu and Zn dopants, beneficial behavior is observed when dopant cations are present in the tetrahedron surrounding the vacancy, although in these cases the effect is more pronounced. In addition, comparison of the vacancy formation energies at different dopant concentrations shows that also the dopant concentration (annex vacancy concentration) has a strong influence on E$_{vac}$. For Cu doping an increase with concentration is shown, while a decrease is seen for both Zn and Gd. The origin of this different behavior may be either the dopant species or the fact that the Cu system contains a too low vacancy concentration per dopant. In the latter case, increasing the Cu concentration also increases the system vacancy concentration. As a result, single oxygen vacancies may interact with different Cu ions, presenting a higher apparent oxygen vacancy concentration for the Cu ions. This reduces the effective vacancy deficiency increasing E$_{vac}$. The same interaction between the vacancies and the dopant elements would, in the case of Zn, result in an apparent over-compensation, or, in the case of Gd, even further increase the already present over-compensation.
\paragraph{Crystal structure.\\}
\indent Where the introduction of a homogeneous distribution of dopants mainly results in an isotropic lattice expansion, the addition of charge compensating oxygen vacancies also results in an increase of the angles between the lattice vectors. Although these changes tend to be quite small ($<5^{\circ}$ in \ce{Ce_{0.75}Cu_{0.25}O_{1.75}}, and $<0.5^{\circ}$ in \ce{Ce_{0.75}Gd_{0.25}O_{1.75}}), they are often anisotropic. As a result we define the change in the lattice parameter for doped systems containing oxygen vacancies as:
\begin{equation}
\Delta a_0 = \frac{(\sqrt[3]{V}-a_{CeO_2})}{a_{CeO_2}} \cdot 100 \%
\end{equation}
with $a_{CeO_2}$ the lattice parameter of pure \ce{CeO2} and $V$ the volume per formula unit of the doped system. Table~\ref{table:MetalXsubst_vac_Ef_BM_dV_da} shows both the change in the volume and lattice parameter for the Cu, Zn, and Gd doped systems. In each case, the oxygen vacancies result in an expansion of the volume (lattice parameter) compared to the system without vacancies, either compensating the lattice compression (Cu and Zn) or further increasing the lattice expansion.\\
\indent In experiments, Bera \textit{et al.} observe only a very small lattice contraction of $-0.01$\% for a system with $5$\% Cu doping.\cite{BeraP:ChemMater2002} This is much less than reported here, but can easily be understood. Firstly, the Cu doped systems presented here, contain much higher Cu concentrations ($25$ and $12.5$\%) than the system of Bera \textit{et al.}, and secondly, the oxygen vacancy concentration in the system of Bera \textit{et al.} contains a much higher relative oxygen vacancy concentration than in the presented systems. As a result, the lattice contraction theoretically presented in Table~\ref{table:MetalXsubst_vac_Ef_BM_dV_da} would be even further compensated if a higher vacancy concentration was used, indicating that values of the order presented by Bera \textit{et al.} are reasonable (though small expansions should also be considered possible).\\
\indent Comparison of $\Delta V$ for \ce{Ce_{0.75}Gd_{0.25}O_{1.75}} in Table~\ref{table:MetalXsubst_vac_Ef_BM_dV_da} ($+1.563$\%) to the values for \ce{Ce_{0.75}Gd_{0.25}O_{1.875}} in Table~\ref{table:MetalXsubst_Gd05Vac_Ef} ($+0.9$\%) shows a clear dependence on the vacancy concentration. As a result, doped \ce{CeO2} compounds should be expected to show breathing behavior under varying oxidizing atmosphere, such as for example car exhaust catalysts.\\
\indent As would be expected from the vacancy induced \ce{Ce^{IV} -> Ce^{III}} transition, Table~\ref{table:MetalXsubst_vac_Ef_BM_dV_da} also shows the volume (lattice parameter) to increase with the number of Ce atoms in the tetrahedral surrounding. This is also in line with earlier results obtained for \ce{Ce_{0.5}La_{0.5}O_{1.75}}.\cite{VanpouckeDannyEP:2011PhysRevB_LCO}
\paragraph{Bulk modulus.\\}
\indent In Sec.~\ref{CeO2pX:sscandCe_vac} it was shown that the introduction of vacancies has a strong influence on the \ce{CeO2} BM. Unlike the volume and lattice parameter change, dopants and the oxygen vacancies have a compound effect on the BM (and TEC), as is seen in Table~\ref{table:MetalXsubst_vac_Ef_BM_dV_da}. It is also interesting to note that the chemical environment of the vacancy has only limited influence on the BM (compare the V$_{A}$ and V$_{B}$ values of the p$222$ supercell), when no charge over-compensation is present. Figure~\ref{fig:TECplotVac}b shows that the decrease of the BM goes hand in hand with the increase of the TEC as was observed for systems without vacancies, showing this behavior to be a universal trend.
\section{Conclusion}\label{CeO2pX:sc_Conclusion}
\indent In summary, we have studied fluorite \ce{CeO2} doped with several aliovalent dopants using \textit{ab-initio} DFT calculations. Dopant concentrations in the range of $0\leq x\leq 25$ \% are investigated, and for Cu, Zn, and Gd dopants also the influence of additional oxygen vacancies is studied.\\
\indent We have shown that for fluorite \ce{CeO2} doped with an aliovalent dopant the lattice expansion shows Vegard law behavior under oxidizing atmosphere. In addition, the Shannon crystal radius of the dopant element can be simply calculated from the lattice parameter, indicating lowered coordination for (most) aliovalent dopants. The introduction of charge compensating oxygen vacancies results in an increase of the lattice parameter, which (partially) compensates the lattice contraction observed for small dopants.\\
\indent As was previously found for group IV dopants, aliovalent dopants also show an inverse relation between the change in bulk modulus and thermal expansion coefficient. Different dopants give rise to different changes in the bulk moduli and thermal expansion coefficients, however, the introduction of oxygen vacancies has a much larger effect, and decreases the bulk modulus
significantly.\\
\indent Defect formation energies are calculated and compared to the oxygen vacancy formation energy to indicate the preference for bulk doping over segregation of the dopant. For the systems investigated we conclude that bulk (substitutional) doping is very unfavorable for Cu, Co, and Zn, while La, Gd, and Sm present themselves as very favorable bulk dopants. No clear relation between the defect formation energy and either the covalent or calculated crystal radius appears to exist.\\
\indent Vacancy formation energies are calculated for different configurations containing $25$ and $12.5$\% Cu, Zn or Gd. For systems where the oxygen vacancies over compensate the charge deficiency due to the aliovalent dopant, the oxygen vacancies are found to be unstable, while being stable otherwise. Although oxygen vacancies are found to stabilize the systems, their contribution remains too small to make bulk doping favorable for Cu, Co and Zn.
\section{Acknowledgement}
\indent The research was financially supported by FWO-Vlaanderen, project n$^{\circ}$ G. $0802.09$N. We also acknowledge the Research Board of the Ghent University. All calculations were carried out using the Stevin Supercomputer Infrastructure at Ghent University.


\bibliography{danny,notes}

\begin{thebibliography}{110}
\expandafter\ifx\csname natexlab\endcsname\relax\def\natexlab#1{#1}\fi
\expandafter\ifx\csname bibnamefont\endcsname\relax
  \def\bibnamefont#1{#1}\fi
\expandafter\ifx\csname bibfnamefont\endcsname\relax
  \def\bibfnamefont#1{#1}\fi
\expandafter\ifx\csname citenamefont\endcsname\relax
  \def\citenamefont#1{#1}\fi
\expandafter\ifx\csname url\endcsname\relax
  \def\url#1{\texttt{#1}}\fi
\expandafter\ifx\csname urlprefix\endcsname\relax\def\urlprefix{URL }\fi
\providecommand{\bibinfo}[2]{#2}
\providecommand{\eprint}[2][]{\url{#2}}

\bibitem[{\citenamefont{Tuller and Nowick}(1975)}]{TullerHL:JES1975}
\bibinfo{author}{\bibfnamefont{H.~L.} \bibnamefont{Tuller}} \bibnamefont{and}
  \bibinfo{author}{\bibfnamefont{A.~S.} \bibnamefont{Nowick}},
  \bibinfo{journal}{J. Electrochem. Soc.} \textbf{\bibinfo{volume}{122}},
  \bibinfo{pages}{255} (\bibinfo{year}{1975}).

\bibitem[{\citenamefont{Trovarelli}(1996)}]{TrovarelliA:CatalRev1996}
\bibinfo{author}{\bibfnamefont{A.}~\bibnamefont{Trovarelli}},
  \bibinfo{journal}{Catal. Rev.-Sci. Eng.} \textbf{\bibinfo{volume}{38}},
  \bibinfo{pages}{439} (\bibinfo{year}{1996}), ISSN \bibinfo{issn}{0161-4940}.

\bibitem[{\citenamefont{Manzoli et~al.}(2008)\citenamefont{Manzoli,
  Avgouropoulos, Tabakova, Papavasiliou, Ioannides, and
  Boccuzzi}}]{ManzoliMaela:CT2008}
\bibinfo{author}{\bibfnamefont{M.}~\bibnamefont{Manzoli}},
  \bibinfo{author}{\bibfnamefont{G.}~\bibnamefont{Avgouropoulos}},
  \bibinfo{author}{\bibfnamefont{T.}~\bibnamefont{Tabakova}},
  \bibinfo{author}{\bibfnamefont{J.}~\bibnamefont{Papavasiliou}},
  \bibinfo{author}{\bibfnamefont{T.}~\bibnamefont{Ioannides}},
  \bibnamefont{and} \bibinfo{author}{\bibfnamefont{F.}~\bibnamefont{Boccuzzi}},
  \bibinfo{journal}{Catal. Today} \textbf{\bibinfo{volume}{138}},
  \bibinfo{pages}{239 } (\bibinfo{year}{2008}).

\bibitem[{\citenamefont{Ka$\check{\mathrm{s}}$par
  et~al.}(1999)\citenamefont{Ka$\check{\mathrm{s}}$par, Fornasiero, and
  Graziani}}]{KasparJ:CatalToday1999}
\bibinfo{author}{\bibfnamefont{J.}~\bibnamefont{Ka$\check{\mathrm{s}}$par}},
  \bibinfo{author}{\bibfnamefont{P.}~\bibnamefont{Fornasiero}},
  \bibnamefont{and} \bibinfo{author}{\bibfnamefont{M.}~\bibnamefont{Graziani}},
  \bibinfo{journal}{Catal. Today} \textbf{\bibinfo{volume}{50}},
  \bibinfo{pages}{285} (\bibinfo{year}{1999}), ISSN \bibinfo{issn}{0920-5861}.

\bibitem[{\citenamefont{Steele}(2000)}]{SteeleBCH:SolStateIon2000}
\bibinfo{author}{\bibfnamefont{B.}~\bibnamefont{Steele}},
  \bibinfo{journal}{Solid State Ionics} \textbf{\bibinfo{volume}{129}},
  \bibinfo{pages}{95} (\bibinfo{year}{2000}), ISSN \bibinfo{issn}{0167-2738},
  \urlprefix\url{http://www.sciencedirect.com/science/article/pii/S0167273899003197}.

\bibitem[{\citenamefont{Shao and Haile}(2004)}]{ShaoZ:nature2004}
\bibinfo{author}{\bibfnamefont{Z.}~\bibnamefont{Shao}} \bibnamefont{and}
  \bibinfo{author}{\bibfnamefont{S.~M.} \bibnamefont{Haile}},
  \bibinfo{journal}{Nature} \textbf{\bibinfo{volume}{431}},
  \bibinfo{pages}{170} (\bibinfo{year}{2004}).

\bibitem[{\citenamefont{Yang et~al.}(2007)\citenamefont{Yang, Luo, Lu, and
  Hermansson}}]{YangZ:JChemPhys2007}
\bibinfo{author}{\bibfnamefont{Z.}~\bibnamefont{Yang}},
  \bibinfo{author}{\bibfnamefont{G.}~\bibnamefont{Luo}},
  \bibinfo{author}{\bibfnamefont{Z.}~\bibnamefont{Lu}}, \bibnamefont{and}
  \bibinfo{author}{\bibfnamefont{K.}~\bibnamefont{Hermansson}},
  \bibinfo{journal}{J. Chem. Phys.} \textbf{\bibinfo{volume}{127}},
  \bibinfo{pages}{074704} (\bibinfo{year}{2007}).

\bibitem[{\citenamefont{Shapovalov and Metiu}(2007)}]{ShapovalovV:JCatal2007}
\bibinfo{author}{\bibfnamefont{V.}~\bibnamefont{Shapovalov}} \bibnamefont{and}
  \bibinfo{author}{\bibfnamefont{H.}~\bibnamefont{Metiu}}, \bibinfo{journal}{J.
  Catal.} \textbf{\bibinfo{volume}{245}}, \bibinfo{pages}{205}
  (\bibinfo{year}{2007}), ISSN \bibinfo{issn}{0021-9517}.

\bibitem[{\citenamefont{Mayernick and
  Janik}(2008)}]{MayernickAD:JPhysChemC2008}
\bibinfo{author}{\bibfnamefont{A.~D.} \bibnamefont{Mayernick}}
  \bibnamefont{and} \bibinfo{author}{\bibfnamefont{M.~J.} \bibnamefont{Janik}},
  \bibinfo{journal}{J. Phys. Chem. C} \textbf{\bibinfo{volume}{112}},
  \bibinfo{pages}{14955} (\bibinfo{year}{2008}).

\bibitem[{\citenamefont{D$\acute{e}$saunay
  et~al.}(2012)\citenamefont{D$\acute{e}$saunay, Ringued$\acute{e}$, Cassir,
  Labat, and Adamo}}]{DesaunayT:SurfSci2012}
\bibinfo{author}{\bibfnamefont{T.}~\bibnamefont{D$\acute{e}$saunay}},
  \bibinfo{author}{\bibfnamefont{A.}~\bibnamefont{Ringued$\acute{e}$}},
  \bibinfo{author}{\bibfnamefont{M.}~\bibnamefont{Cassir}},
  \bibinfo{author}{\bibfnamefont{F.}~\bibnamefont{Labat}}, \bibnamefont{and}
  \bibinfo{author}{\bibfnamefont{C.}~\bibnamefont{Adamo}},
  \bibinfo{journal}{Surf. Sci.} \textbf{\bibinfo{volume}{606}},
  \bibinfo{pages}{305} (\bibinfo{year}{2012}), ISSN \bibinfo{issn}{0039-6028}.

\bibitem[{\citenamefont{Yao and Yu~Yao}(1984)}]{YaoHC:JCatal1984}
\bibinfo{author}{\bibfnamefont{H.}~\bibnamefont{Yao}} \bibnamefont{and}
  \bibinfo{author}{\bibfnamefont{Y.}~\bibnamefont{Yu~Yao}},
  \bibinfo{journal}{J. Catal.} \textbf{\bibinfo{volume}{86}},
  \bibinfo{pages}{254} (\bibinfo{year}{1984}), ISSN \bibinfo{issn}{0021-9517},
  \urlprefix\url{http://www.sciencedirect.com/science/article/pii/0021951784903713}.

\bibitem[{\citenamefont{Fu et~al.}(2003)\citenamefont{Fu, Saltsburg, and
  Flytzani-Stephanopoulos}}]{FuQ:2003Science}
\bibinfo{author}{\bibfnamefont{Q.}~\bibnamefont{Fu}},
  \bibinfo{author}{\bibfnamefont{H.}~\bibnamefont{Saltsburg}},
  \bibnamefont{and}
  \bibinfo{author}{\bibfnamefont{M.}~\bibnamefont{Flytzani-Stephanopoulos}},
  \bibinfo{journal}{Science} \textbf{\bibinfo{volume}{301}},
  \bibinfo{pages}{935} (\bibinfo{year}{2003}),
  \eprint{http://www.sciencemag.org/content/301/5635/935.full.pdf},
  \urlprefix\url{http://www.sciencemag.org/content/301/5635/935.abstract}.

\bibitem[{\citenamefont{McBride et~al.}(1994)\citenamefont{McBride, Hass,
  Poindexter, and Weber}}]{McBrideJR:JApplPhys1994}
\bibinfo{author}{\bibfnamefont{J.~R.} \bibnamefont{McBride}},
  \bibinfo{author}{\bibfnamefont{K.~C.} \bibnamefont{Hass}},
  \bibinfo{author}{\bibfnamefont{B.~D.} \bibnamefont{Poindexter}},
  \bibnamefont{and} \bibinfo{author}{\bibfnamefont{W.~H.} \bibnamefont{Weber}},
  \bibinfo{journal}{J. Appl. Phys.} \textbf{\bibinfo{volume}{76}},
  \bibinfo{pages}{2435} (\bibinfo{year}{1994}),
  \urlprefix\url{http://link.aip.org/link/?JAP/76/2435/1}.

\bibitem[{\citenamefont{Cao et~al.}(2003)\citenamefont{Cao, Vassen, Fischer,
  Tietz, Jungen, and Stover}}]{CaoXueqiang:AdvMat2003}
\bibinfo{author}{\bibfnamefont{X.}~\bibnamefont{Cao}},
  \bibinfo{author}{\bibfnamefont{R.}~\bibnamefont{Vassen}},
  \bibinfo{author}{\bibfnamefont{W.}~\bibnamefont{Fischer}},
  \bibinfo{author}{\bibfnamefont{F.}~\bibnamefont{Tietz}},
  \bibinfo{author}{\bibfnamefont{W.}~\bibnamefont{Jungen}}, \bibnamefont{and}
  \bibinfo{author}{\bibfnamefont{D.}~\bibnamefont{Stover}},
  \bibinfo{journal}{Adv. Mater.} \textbf{\bibinfo{volume}{15}},
  \bibinfo{pages}{1438} (\bibinfo{year}{2003}).

\bibitem[{\citenamefont{Cao et~al.}(2004)\citenamefont{Cao, Vassen, and
  Stoever}}]{CaoXueqiang:JEurCeramSoc2004}
\bibinfo{author}{\bibfnamefont{X.~Q.} \bibnamefont{Cao}},
  \bibinfo{author}{\bibfnamefont{R.}~\bibnamefont{Vassen}}, \bibnamefont{and}
  \bibinfo{author}{\bibfnamefont{D.}~\bibnamefont{Stoever}},
  \bibinfo{journal}{J. Eur. Ceram. Soc.} \textbf{\bibinfo{volume}{24}},
  \bibinfo{pages}{1} (\bibinfo{year}{2004}).

\bibitem[{\citenamefont{Paranthaman et~al.}(1997)\citenamefont{Paranthaman,
  Goyal, List, Specht, Lee, Martin, He, Christen, Norton, Budai
  et~al.}}]{ParanthamanM:1997PhysC}
\bibinfo{author}{\bibfnamefont{M.}~\bibnamefont{Paranthaman}},
  \bibinfo{author}{\bibfnamefont{A.}~\bibnamefont{Goyal}},
  \bibinfo{author}{\bibfnamefont{F.}~\bibnamefont{List}},
  \bibinfo{author}{\bibfnamefont{E.}~\bibnamefont{Specht}},
  \bibinfo{author}{\bibfnamefont{D.}~\bibnamefont{Lee}},
  \bibinfo{author}{\bibfnamefont{P.}~\bibnamefont{Martin}},
  \bibinfo{author}{\bibfnamefont{Q.}~\bibnamefont{He}},
  \bibinfo{author}{\bibfnamefont{D.}~\bibnamefont{Christen}},
  \bibinfo{author}{\bibfnamefont{D.}~\bibnamefont{Norton}},
  \bibinfo{author}{\bibfnamefont{J.}~\bibnamefont{Budai}},
  \bibnamefont{et~al.}, \bibinfo{journal}{Physica C}
  \textbf{\bibinfo{volume}{275}}, \bibinfo{pages}{266 } (\bibinfo{year}{1997}).

\bibitem[{\citenamefont{Oh et~al.}(1998)\citenamefont{Oh, Yoo, Lee, Kim, and
  Youm}}]{OhSanghyun:PhysC1998}
\bibinfo{author}{\bibfnamefont{S.}~\bibnamefont{Oh}},
  \bibinfo{author}{\bibfnamefont{J.}~\bibnamefont{Yoo}},
  \bibinfo{author}{\bibfnamefont{K.}~\bibnamefont{Lee}},
  \bibinfo{author}{\bibfnamefont{J.}~\bibnamefont{Kim}}, \bibnamefont{and}
  \bibinfo{author}{\bibfnamefont{D.}~\bibnamefont{Youm}},
  \bibinfo{journal}{Physica C} \textbf{\bibinfo{volume}{308}},
  \bibinfo{pages}{91 } (\bibinfo{year}{1998}).

\bibitem[{\citenamefont{Penneman et~al.}(2004)\citenamefont{Penneman,
  Van~Driessche, Bruneel, and Hoste}}]{PennemanG:2004EuroCeram}
\bibinfo{author}{\bibfnamefont{G.}~\bibnamefont{Penneman}},
  \bibinfo{author}{\bibfnamefont{I.}~\bibnamefont{Van~Driessche}},
  \bibinfo{author}{\bibfnamefont{E.}~\bibnamefont{Bruneel}}, \bibnamefont{and}
  \bibinfo{author}{\bibfnamefont{S.}~\bibnamefont{Hoste}}, in
  \emph{\bibinfo{booktitle}{{Euro Ceramics VIII, Pts 1-3}}}, edited by
  \bibinfo{editor}{\bibnamefont{{Mandal, H and Ovecoglu, L}}}
  (\bibinfo{organization}{{Turkish Ceram Soc; European Ceram Soc}},
  \bibinfo{year}{2004}), vol. \bibinfo{volume}{264--268} of
  \emph{\bibinfo{series}{{Key Engineering Materials}}}, pp.
  \bibinfo{pages}{501--504}, ISBN \bibinfo{isbn}{0-87849-946-6}, ISSN
  \bibinfo{issn}{1013-9826}, \bibinfo{note}{{8th Conference of the
  European-Ceramic-Society, Istanbul, Turkey, jun 29-jul 03, 2003}}.

\bibitem[{\citenamefont{Takahashi et~al.}(2004)\citenamefont{Takahashi, Aoki,
  Hasegawa, Maeda, Honjo, Yamada, and Shiohara}}]{TakahashiY:PhysC2004}
\bibinfo{author}{\bibfnamefont{Y.}~\bibnamefont{Takahashi}},
  \bibinfo{author}{\bibfnamefont{Y.}~\bibnamefont{Aoki}},
  \bibinfo{author}{\bibfnamefont{T.}~\bibnamefont{Hasegawa}},
  \bibinfo{author}{\bibfnamefont{T.}~\bibnamefont{Maeda}},
  \bibinfo{author}{\bibfnamefont{T.}~\bibnamefont{Honjo}},
  \bibinfo{author}{\bibfnamefont{Y.}~\bibnamefont{Yamada}}, \bibnamefont{and}
  \bibinfo{author}{\bibfnamefont{Y.}~\bibnamefont{Shiohara}},
  \bibinfo{journal}{Physica C} \textbf{\bibinfo{volume}{412-414, Part 2}},
  \bibinfo{pages}{905} (\bibinfo{year}{2004}).

\bibitem[{\citenamefont{Knoth et~al.}(2005)\citenamefont{Knoth, Schlobach,
  H{\"u}hne, Schultz, and Holzapfel}}]{KnothKerstin:PhysC2005}
\bibinfo{author}{\bibfnamefont{K.}~\bibnamefont{Knoth}},
  \bibinfo{author}{\bibfnamefont{B.}~\bibnamefont{Schlobach}},
  \bibinfo{author}{\bibfnamefont{R.}~\bibnamefont{H{\"u}hne}},
  \bibinfo{author}{\bibfnamefont{L.}~\bibnamefont{Schultz}}, \bibnamefont{and}
  \bibinfo{author}{\bibfnamefont{B.}~\bibnamefont{Holzapfel}},
  \bibinfo{journal}{Physica C} \textbf{\bibinfo{volume}{426-431, Part 2}},
  \bibinfo{pages}{979} (\bibinfo{year}{2005}).

\bibitem[{\citenamefont{Van~de Velde et~al.}(2010)\citenamefont{Van~de Velde,
  Van~de Vyver, Brunkahl, Hoste, Bruneel, and
  Van~Driessche}}]{VandeVeldeNigel:EurJInorChem2010}
\bibinfo{author}{\bibfnamefont{N.}~\bibnamefont{Van~de Velde}},
  \bibinfo{author}{\bibfnamefont{D.}~\bibnamefont{Van~de Vyver}},
  \bibinfo{author}{\bibfnamefont{O.}~\bibnamefont{Brunkahl}},
  \bibinfo{author}{\bibfnamefont{S.}~\bibnamefont{Hoste}},
  \bibinfo{author}{\bibfnamefont{E.}~\bibnamefont{Bruneel}}, \bibnamefont{and}
  \bibinfo{author}{\bibfnamefont{I.}~\bibnamefont{Van~Driessche}},
  \bibinfo{journal}{Eur. J. Inor. Chem.} pp. \bibinfo{pages}{233--241}
  (\bibinfo{year}{2010}).

\bibitem[{\citenamefont{Narayanan et~al.}(2012)\citenamefont{Narayanan,
  Lommens, De~Buysser, Vanpoucke, Huehne, Molina, Van~Tendeloo, Van Der~Voort,
  and Van~Driessche}}]{VyshnaviN:2012JMaterChem}
\bibinfo{author}{\bibfnamefont{V.}~\bibnamefont{Narayanan}},
  \bibinfo{author}{\bibfnamefont{P.}~\bibnamefont{Lommens}},
  \bibinfo{author}{\bibfnamefont{K.}~\bibnamefont{De~Buysser}},
  \bibinfo{author}{\bibfnamefont{D.~E.~P.} \bibnamefont{Vanpoucke}},
  \bibinfo{author}{\bibfnamefont{R.}~\bibnamefont{Huehne}},
  \bibinfo{author}{\bibfnamefont{L.}~\bibnamefont{Molina}},
  \bibinfo{author}{\bibfnamefont{G.}~\bibnamefont{Van~Tendeloo}},
  \bibinfo{author}{\bibfnamefont{P.}~\bibnamefont{Van Der~Voort}},
  \bibnamefont{and}
  \bibinfo{author}{\bibfnamefont{I.}~\bibnamefont{Van~Driessche}},
  \bibinfo{journal}{J. Mater. Chem.} \textbf{\bibinfo{volume}{22}},
  \bibinfo{pages}{8476} (\bibinfo{year}{2012}),
  \urlprefix\url{http://dx.doi.org/10.1039/C2JM15752G}.

\bibitem[{\citenamefont{Mogensen et~al.}(2000)\citenamefont{Mogensen, Sammes,
  and Tompsett}}]{MogensenM:SolStateI2000}
\bibinfo{author}{\bibfnamefont{M.}~\bibnamefont{Mogensen}},
  \bibinfo{author}{\bibfnamefont{N.~M.} \bibnamefont{Sammes}},
  \bibnamefont{and} \bibinfo{author}{\bibfnamefont{G.~A.}
  \bibnamefont{Tompsett}}, \bibinfo{journal}{Solid State Ionics}
  \textbf{\bibinfo{volume}{129}}, \bibinfo{pages}{63} (\bibinfo{year}{2000}),
  ISSN \bibinfo{issn}{0167-2738},
  \urlprefix\url{http://www.sciencedirect.com/science/article/pii/S0167273899003185}.

\bibitem[{\citenamefont{Bera et~al.}(2002)\citenamefont{Bera, Priolkar, Sarode,
  Hegde, Emura, Kumashiro, and Lalla}}]{BeraP:ChemMater2002}
\bibinfo{author}{\bibfnamefont{P.}~\bibnamefont{Bera}},
  \bibinfo{author}{\bibfnamefont{K.~R.} \bibnamefont{Priolkar}},
  \bibinfo{author}{\bibfnamefont{P.~R.} \bibnamefont{Sarode}},
  \bibinfo{author}{\bibfnamefont{M.~S.} \bibnamefont{Hegde}},
  \bibinfo{author}{\bibfnamefont{S.}~\bibnamefont{Emura}},
  \bibinfo{author}{\bibfnamefont{R.}~\bibnamefont{Kumashiro}},
  \bibnamefont{and} \bibinfo{author}{\bibfnamefont{N.~P.} \bibnamefont{Lalla}},
  \bibinfo{journal}{Chem. Mater.} \textbf{\bibinfo{volume}{14}},
  \bibinfo{pages}{3591} (\bibinfo{year}{2002}),
  \eprint{http://pubs.acs.org/doi/pdf/10.1021/cm0201706},
  \urlprefix\url{http://pubs.acs.org/doi/abs/10.1021/cm0201706}.

\bibitem[{\citenamefont{Li et~al.}(2008)\citenamefont{Li, Wei, and
  Pan}}]{LiB:JPowerSources2008}
\bibinfo{author}{\bibfnamefont{B.}~\bibnamefont{Li}},
  \bibinfo{author}{\bibfnamefont{X.}~\bibnamefont{Wei}}, \bibnamefont{and}
  \bibinfo{author}{\bibfnamefont{W.}~\bibnamefont{Pan}}, \bibinfo{journal}{J.
  Power Sources} \textbf{\bibinfo{volume}{183}}, \bibinfo{pages}{498}
  (\bibinfo{year}{2008}), ISSN \bibinfo{issn}{0378-7753}.

\bibitem[{\citenamefont{Mullins et~al.}(1999)\citenamefont{Mullins, Radulovic,
  and Overbury}}]{MullinsDR:SurfSci1999}
\bibinfo{author}{\bibfnamefont{D.}~\bibnamefont{Mullins}},
  \bibinfo{author}{\bibfnamefont{P.}~\bibnamefont{Radulovic}},
  \bibnamefont{and} \bibinfo{author}{\bibfnamefont{S.}~\bibnamefont{Overbury}},
  \bibinfo{journal}{Surf. Sci.} \textbf{\bibinfo{volume}{429}},
  \bibinfo{pages}{186} (\bibinfo{year}{1999}), ISSN \bibinfo{issn}{0039-6028}.

\bibitem[{\citenamefont{Aspinall et~al.}(2011)\citenamefont{Aspinall, Bacsa,
  Jones, Wrench, Black, Chalker, King, Marshall, Werner, Davies
  et~al.}}]{AspinallHC:InorgChem2011}
\bibinfo{author}{\bibfnamefont{H.~C.} \bibnamefont{Aspinall}},
  \bibinfo{author}{\bibfnamefont{J.}~\bibnamefont{Bacsa}},
  \bibinfo{author}{\bibfnamefont{A.~C.} \bibnamefont{Jones}},
  \bibinfo{author}{\bibfnamefont{J.~S.} \bibnamefont{Wrench}},
  \bibinfo{author}{\bibfnamefont{K.}~\bibnamefont{Black}},
  \bibinfo{author}{\bibfnamefont{P.~R.} \bibnamefont{Chalker}},
  \bibinfo{author}{\bibfnamefont{P.~J.} \bibnamefont{King}},
  \bibinfo{author}{\bibfnamefont{P.}~\bibnamefont{Marshall}},
  \bibinfo{author}{\bibfnamefont{M.}~\bibnamefont{Werner}},
  \bibinfo{author}{\bibfnamefont{H.~O.} \bibnamefont{Davies}},
  \bibnamefont{et~al.}, \bibinfo{journal}{Inorg. Chem.}
  \textbf{\bibinfo{volume}{50}}, \bibinfo{pages}{11644} (\bibinfo{year}{2011}),
  \urlprefix\url{http://pubs.acs.org/doi/abs/10.1021/ic201593s}.

\bibitem[{\citenamefont{Rossignol et~al.}(1999)\citenamefont{Rossignol, Gerard,
  and Duprez}}]{RossignolS:JMaterChem1999}
\bibinfo{author}{\bibfnamefont{S.}~\bibnamefont{Rossignol}},
  \bibinfo{author}{\bibfnamefont{F.}~\bibnamefont{Gerard}}, \bibnamefont{and}
  \bibinfo{author}{\bibfnamefont{D.}~\bibnamefont{Duprez}},
  \bibinfo{journal}{J. Mater. Chem.} \textbf{\bibinfo{volume}{9}},
  \bibinfo{pages}{1615} (\bibinfo{year}{1999}),
  \urlprefix\url{http://dx.doi.org/10.1039/A900536F}.

\bibitem[{\citenamefont{Van~Driessche
  et~al.}({2002})\citenamefont{Van~Driessche, Penneman, De~Meyer, Stambolova,
  Bruneel, and Hoste}}]{VanDriesscheI:2002ECVII}
\bibinfo{author}{\bibfnamefont{I.}~\bibnamefont{Van~Driessche}},
  \bibinfo{author}{\bibfnamefont{G.}~\bibnamefont{Penneman}},
  \bibinfo{author}{\bibfnamefont{C.}~\bibnamefont{De~Meyer}},
  \bibinfo{author}{\bibfnamefont{I.}~\bibnamefont{Stambolova}},
  \bibinfo{author}{\bibfnamefont{E.}~\bibnamefont{Bruneel}}, \bibnamefont{and}
  \bibinfo{author}{\bibfnamefont{S.}~\bibnamefont{Hoste}}, in
  \emph{\bibinfo{booktitle}{{Euro Ceramics VII, PT 1-3}}}
  (\bibinfo{publisher}{{Trans Tech Publications Ltd}},
  \bibinfo{address}{{Brandrain 6, CH-8707 Zurich-Uetikon, Switzerland}},
  \bibinfo{year}{{2002}}), vol. \bibinfo{volume}{{206-2}} of
  \emph{\bibinfo{series}{{Key Engineering Materials}}}, pp.
  \bibinfo{pages}{{479--482}}.

\bibitem[{\citenamefont{Claparede et~al.}(2011)\citenamefont{Claparede,
  Clavier, Dacheux, Moisy, Podor, and Ravaux}}]{ClaparedeL:InorChem2011}
\bibinfo{author}{\bibfnamefont{L.}~\bibnamefont{Claparede}},
  \bibinfo{author}{\bibfnamefont{N.}~\bibnamefont{Clavier}},
  \bibinfo{author}{\bibfnamefont{N.}~\bibnamefont{Dacheux}},
  \bibinfo{author}{\bibfnamefont{P.}~\bibnamefont{Moisy}},
  \bibinfo{author}{\bibfnamefont{R.}~\bibnamefont{Podor}}, \bibnamefont{and}
  \bibinfo{author}{\bibfnamefont{J.}~\bibnamefont{Ravaux}},
  \bibinfo{journal}{Inor. Chem.} \textbf{\bibinfo{volume}{50}},
  \bibinfo{pages}{9059} (\bibinfo{year}{2011}),
  \eprint{http://pubs.acs.org/doi/pdf/10.1021/ic201269c},
  \urlprefix\url{http://pubs.acs.org/doi/abs/10.1021/ic201269c}.

\bibitem[{\citenamefont{Horlait et~al.}(2012)\citenamefont{Horlait, Clavier,
  Szenknect, Dacheux, and Dubois}}]{HorlaitD:InorChem2012}
\bibinfo{author}{\bibfnamefont{D.}~\bibnamefont{Horlait}},
  \bibinfo{author}{\bibfnamefont{N.}~\bibnamefont{Clavier}},
  \bibinfo{author}{\bibfnamefont{S.}~\bibnamefont{Szenknect}},
  \bibinfo{author}{\bibfnamefont{N.}~\bibnamefont{Dacheux}}, \bibnamefont{and}
  \bibinfo{author}{\bibfnamefont{V.}~\bibnamefont{Dubois}},
  \bibinfo{journal}{Inor. Chem.} \textbf{\bibinfo{volume}{51}},
  \bibinfo{pages}{3868} (\bibinfo{year}{2012}),
  \eprint{http://pubs.acs.org/doi/pdf/10.1021/ic300071c},
  \urlprefix\url{http://pubs.acs.org/doi/abs/10.1021/ic300071c}.

\bibitem[{\citenamefont{Song et~al.}(2009)\citenamefont{Song, Zhang, Yang, Liu,
  Li, Shah, Zhu, and Xiao}}]{SongYQ:JPhysCondensMatter2009}
\bibinfo{author}{\bibfnamefont{Y.~Q.} \bibnamefont{Song}},
  \bibinfo{author}{\bibfnamefont{H.~W.} \bibnamefont{Zhang}},
  \bibinfo{author}{\bibfnamefont{Q.~H.} \bibnamefont{Yang}},
  \bibinfo{author}{\bibfnamefont{Y.~L.} \bibnamefont{Liu}},
  \bibinfo{author}{\bibfnamefont{Y.~X.} \bibnamefont{Li}},
  \bibinfo{author}{\bibfnamefont{L.~R.} \bibnamefont{Shah}},
  \bibinfo{author}{\bibfnamefont{H.}~\bibnamefont{Zhu}}, \bibnamefont{and}
  \bibinfo{author}{\bibfnamefont{J.~Q.} \bibnamefont{Xiao}},
  \bibinfo{journal}{J. Phys.: Condens. Matter} \textbf{\bibinfo{volume}{21}},
  \bibinfo{pages}{125504} (\bibinfo{year}{2009}),
  \urlprefix\url{http://stacks.iop.org/0953-8984/21/i=12/a=125504}.

\bibitem[{\citenamefont{Yang et~al.}(2008)\citenamefont{Yang, Wei, Fu, Lu, and
  Hermansson}}]{YangZ:SurfSci2008}
\bibinfo{author}{\bibfnamefont{Z.}~\bibnamefont{Yang}},
  \bibinfo{author}{\bibfnamefont{Y.}~\bibnamefont{Wei}},
  \bibinfo{author}{\bibfnamefont{Z.}~\bibnamefont{Fu}},
  \bibinfo{author}{\bibfnamefont{Z.}~\bibnamefont{Lu}}, \bibnamefont{and}
  \bibinfo{author}{\bibfnamefont{K.}~\bibnamefont{Hermansson}},
  \bibinfo{journal}{Surf. Sci.} \textbf{\bibinfo{volume}{602}},
  \bibinfo{pages}{1199} (\bibinfo{year}{2008}), ISSN \bibinfo{issn}{0039-6028}.

\bibitem[{\citenamefont{Andersson et~al.}(2006)\citenamefont{Andersson, Simak,
  Skorodumova, Abrikosov, and Johansson}}]{AnderssonDA:PNAS2006}
\bibinfo{author}{\bibfnamefont{D.~A.} \bibnamefont{Andersson}},
  \bibinfo{author}{\bibfnamefont{S.~I.} \bibnamefont{Simak}},
  \bibinfo{author}{\bibfnamefont{N.~V.} \bibnamefont{Skorodumova}},
  \bibinfo{author}{\bibfnamefont{I.~A.} \bibnamefont{Abrikosov}},
  \bibnamefont{and}
  \bibinfo{author}{\bibfnamefont{B.}~\bibnamefont{Johansson}},
  \bibinfo{journal}{{PNAS}} \textbf{\bibinfo{volume}{103}},
  \bibinfo{pages}{3518} (\bibinfo{year}{2006}).

\bibitem[{\citenamefont{Lu et~al.}(2011)\citenamefont{Lu, Yang, He, Castleton,
  and Hermansson}}]{LuZ:ChemPhysLett2011}
\bibinfo{author}{\bibfnamefont{Z.}~\bibnamefont{Lu}},
  \bibinfo{author}{\bibfnamefont{Z.}~\bibnamefont{Yang}},
  \bibinfo{author}{\bibfnamefont{B.}~\bibnamefont{He}},
  \bibinfo{author}{\bibfnamefont{C.}~\bibnamefont{Castleton}},
  \bibnamefont{and}
  \bibinfo{author}{\bibfnamefont{K.}~\bibnamefont{Hermansson}},
  \bibinfo{journal}{Chem. Phys. Lett.} \textbf{\bibinfo{volume}{510}},
  \bibinfo{pages}{60} (\bibinfo{year}{2011}), ISSN \bibinfo{issn}{0009-2614}.

\bibitem[{\citenamefont{Vanpoucke et~al.}(2011)\citenamefont{Vanpoucke,
  Bultinck, Cottenier, Van~Speybroeck, and
  Van~Driessche}}]{VanpouckeDannyEP:2011PhysRevB_LCO}
\bibinfo{author}{\bibfnamefont{D.~E.~P.} \bibnamefont{Vanpoucke}},
  \bibinfo{author}{\bibfnamefont{P.}~\bibnamefont{Bultinck}},
  \bibinfo{author}{\bibfnamefont{S.}~\bibnamefont{Cottenier}},
  \bibinfo{author}{\bibfnamefont{V.}~\bibnamefont{Van~Speybroeck}},
  \bibnamefont{and}
  \bibinfo{author}{\bibfnamefont{I.}~\bibnamefont{Van~Driessche}},
  \bibinfo{journal}{Phys. Rev. B} \textbf{\bibinfo{volume}{84}},
  \bibinfo{pages}{054110} (\bibinfo{year}{2011}),
  \urlprefix\url{http://link.aps.org/doi/10.1103/PhysRevB.84.054110}.

\bibitem[{\citenamefont{Minervini et~al.}(2000)\citenamefont{Minervini, Grimes,
  and Sickafus}}]{MinerviniL:JAmCeramSoc2000}
\bibinfo{author}{\bibfnamefont{L.}~\bibnamefont{Minervini}},
  \bibinfo{author}{\bibfnamefont{R.~W.} \bibnamefont{Grimes}},
  \bibnamefont{and} \bibinfo{author}{\bibfnamefont{K.~E.}
  \bibnamefont{Sickafus}}, \bibinfo{journal}{J. Am. Ceram. Soc.}
  \textbf{\bibinfo{volume}{83}}, \bibinfo{pages}{1873} (\bibinfo{year}{2000}).

\bibitem[{\citenamefont{Ye et~al.}(2009)\citenamefont{Ye, Mori, Ou, and
  Cormack}}]{YeF:SolStateIon2009}
\bibinfo{author}{\bibfnamefont{F.}~\bibnamefont{Ye}},
  \bibinfo{author}{\bibfnamefont{T.}~\bibnamefont{Mori}},
  \bibinfo{author}{\bibfnamefont{D.~R.} \bibnamefont{Ou}}, \bibnamefont{and}
  \bibinfo{author}{\bibfnamefont{A.~N.} \bibnamefont{Cormack}},
  \bibinfo{journal}{Solid State Ionics} \textbf{\bibinfo{volume}{180}},
  \bibinfo{pages}{1127 } (\bibinfo{year}{2009}), ISSN
  \bibinfo{issn}{0167-2738}.

\bibitem[{\citenamefont{Wei et~al.}(2009)\citenamefont{Wei, Pan, Cheng, and
  Li}}]{WeiX:SolStateIonics2009}
\bibinfo{author}{\bibfnamefont{X.}~\bibnamefont{Wei}},
  \bibinfo{author}{\bibfnamefont{W.}~\bibnamefont{Pan}},
  \bibinfo{author}{\bibfnamefont{L.}~\bibnamefont{Cheng}}, \bibnamefont{and}
  \bibinfo{author}{\bibfnamefont{B.}~\bibnamefont{Li}}, \bibinfo{journal}{Solid
  State Ionics} \textbf{\bibinfo{volume}{180}}, \bibinfo{pages}{13}
  (\bibinfo{year}{2009}), ISSN \bibinfo{issn}{0167-2738}.

\bibitem[{\citenamefont{Ryan et~al.}(2003)\citenamefont{Ryan, McGrath, Farrell,
  O'Neill, Barnes, and Morris}}]{RyanKM:JPCondMat2003}
\bibinfo{author}{\bibfnamefont{K.~M.} \bibnamefont{Ryan}},
  \bibinfo{author}{\bibfnamefont{J.~P.} \bibnamefont{McGrath}},
  \bibinfo{author}{\bibfnamefont{R.~A.} \bibnamefont{Farrell}},
  \bibinfo{author}{\bibfnamefont{W.~M.} \bibnamefont{O'Neill}},
  \bibinfo{author}{\bibfnamefont{C.~J.} \bibnamefont{Barnes}},
  \bibnamefont{and} \bibinfo{author}{\bibfnamefont{M.~A.}
  \bibnamefont{Morris}}, \bibinfo{journal}{J. Phys. Condens. Matter}
  \textbf{\bibinfo{volume}{15}}, \bibinfo{pages}{L49} (\bibinfo{year}{2003}).

\bibitem[{\citenamefont{Bezerra~Lopes et~al.}(2009)\citenamefont{Bezerra~Lopes,
  de~Souza, Vieira~de Morais, Dallas, and
  Gavarri}}]{LopesFWB:Hydrometallurgy2009}
\bibinfo{author}{\bibfnamefont{F.~W.} \bibnamefont{Bezerra~Lopes}},
  \bibinfo{author}{\bibfnamefont{C.~P.} \bibnamefont{de~Souza}},
  \bibinfo{author}{\bibfnamefont{A.~M.} \bibnamefont{Vieira~de Morais}},
  \bibinfo{author}{\bibfnamefont{J.-P.} \bibnamefont{Dallas}},
  \bibnamefont{and} \bibinfo{author}{\bibfnamefont{J.-R.}
  \bibnamefont{Gavarri}}, \bibinfo{journal}{Hydrometallurgy}
  \textbf{\bibinfo{volume}{97}}, \bibinfo{pages}{167} (\bibinfo{year}{2009}).

\bibitem[{\citenamefont{Reddy et~al.}(2010)\citenamefont{Reddy, Katta, and
  Thrimurthulu}}]{ReddyB:2010CM}
\bibinfo{author}{\bibfnamefont{M.}~\bibnamefont{Reddy},
  \bibfnamefont{Benjaram}},
  \bibinfo{author}{\bibfnamefont{L.}~\bibnamefont{Katta}}, \bibnamefont{and}
  \bibinfo{author}{\bibfnamefont{G.}~\bibnamefont{Thrimurthulu}},
  \bibinfo{journal}{Chem. Mater.} \textbf{\bibinfo{volume}{22}},
  \bibinfo{pages}{467} (\bibinfo{year}{2010}).

\bibitem[{\citenamefont{Horlait et~al.}(2011)\citenamefont{Horlait,
  Clapar\`{e}de, Clavier, Szenknect, Dacheux, Ravaux, and
  Podor}}]{HorlaitD:2011InorChem}
\bibinfo{author}{\bibfnamefont{D.}~\bibnamefont{Horlait}},
  \bibinfo{author}{\bibfnamefont{L.}~\bibnamefont{Clapar\`{e}de}},
  \bibinfo{author}{\bibfnamefont{N.}~\bibnamefont{Clavier}},
  \bibinfo{author}{\bibfnamefont{S.}~\bibnamefont{Szenknect}},
  \bibinfo{author}{\bibfnamefont{N.}~\bibnamefont{Dacheux}},
  \bibinfo{author}{\bibfnamefont{J.}~\bibnamefont{Ravaux}}, \bibnamefont{and}
  \bibinfo{author}{\bibfnamefont{R.}~\bibnamefont{Podor}},
  \bibinfo{journal}{Inor. Chem.} \textbf{\bibinfo{volume}{50}},
  \bibinfo{pages}{7150} (\bibinfo{year}{2011}),
  \eprint{http://pubs.acs.org/doi/pdf/10.1021/ic200751m},
  \urlprefix\url{http://pubs.acs.org/doi/abs/10.1021/ic200751m}.

\bibitem[{\citenamefont{Andersson
  et~al.}(2007{\natexlab{a}})\citenamefont{Andersson, Simak, Skorodumova,
  Abrikosov, and Johansson}}]{AnderssonDA:2007bPhysRevB}
\bibinfo{author}{\bibfnamefont{D.~A.} \bibnamefont{Andersson}},
  \bibinfo{author}{\bibfnamefont{S.~I.} \bibnamefont{Simak}},
  \bibinfo{author}{\bibfnamefont{N.~V.} \bibnamefont{Skorodumova}},
  \bibinfo{author}{\bibfnamefont{I.~A.} \bibnamefont{Abrikosov}},
  \bibnamefont{and}
  \bibinfo{author}{\bibfnamefont{B.}~\bibnamefont{Johansson}},
  \bibinfo{journal}{Phys. Rev. B} \textbf{\bibinfo{volume}{76}},
  \bibinfo{pages}{174119} (\bibinfo{year}{2007}{\natexlab{a}}),
  \urlprefix\url{http://link.aps.org/doi/10.1103/PhysRevB.76.174119}.

\bibitem[{\citenamefont{Andersson
  et~al.}(2007{\natexlab{b}})\citenamefont{Andersson, Simak, Skorodumova,
  Abrikosov, and Johansson}}]{AnderssonDA:2007ApplPhysLett}
\bibinfo{author}{\bibfnamefont{D.~A.} \bibnamefont{Andersson}},
  \bibinfo{author}{\bibfnamefont{S.~I.} \bibnamefont{Simak}},
  \bibinfo{author}{\bibfnamefont{N.~V.} \bibnamefont{Skorodumova}},
  \bibinfo{author}{\bibfnamefont{I.~A.} \bibnamefont{Abrikosov}},
  \bibnamefont{and}
  \bibinfo{author}{\bibfnamefont{B.}~\bibnamefont{Johansson}},
  \bibinfo{journal}{Appl. Phys. Lett.} \textbf{\bibinfo{volume}{90}},
  \bibinfo{pages}{031909} (\bibinfo{year}{2007}{\natexlab{b}}),
  \urlprefix\url{http://link.aip.org/link/?APL/90/031909/1}.

\bibitem[{\citenamefont{Vanpoucke et~al.}(2012)\citenamefont{Vanpoucke,
  Cottenier, Van~Speybroeck, Bultinck, and
  Van~Driessche}}]{VanpouckeDannyEP:2012aApplSurfSci}
\bibinfo{author}{\bibfnamefont{D.~E.~P.} \bibnamefont{Vanpoucke}},
  \bibinfo{author}{\bibfnamefont{S.}~\bibnamefont{Cottenier}},
  \bibinfo{author}{\bibfnamefont{V.}~\bibnamefont{Van~Speybroeck}},
  \bibinfo{author}{\bibfnamefont{P.}~\bibnamefont{Bultinck}}, \bibnamefont{and}
  \bibinfo{author}{\bibfnamefont{I.}~\bibnamefont{Van~Driessche}},
  \bibinfo{journal}{Appl. Surf. Sci.} \textbf{\bibinfo{volume}{260}},
  \bibinfo{pages}{32} (\bibinfo{year}{2012}), ISSN \bibinfo{issn}{0169-4332},
  \bibinfo{note}{<ce:title>EMRS 2011 Fall meeting symposium on Stress,
  structure and stoichiometry effects on nanomaterials</ce:title>},
  \urlprefix\url{http://www.sciencedirect.com/science/article/pii/S0169433212000451}.

\bibitem[{\citenamefont{Vanpoucke et~al.}(2013)\citenamefont{Vanpoucke,
  Cottenier, Van~Speybroeck, Van~Driessche, and
  Bultinck}}]{VanpouckeDanny:2012dGroupIVdopants}
\bibinfo{author}{\bibfnamefont{D.~E.~P.} \bibnamefont{Vanpoucke}},
  \bibinfo{author}{\bibfnamefont{S.}~\bibnamefont{Cottenier}},
  \bibinfo{author}{\bibfnamefont{V.}~\bibnamefont{Van~Speybroeck}},
  \bibinfo{author}{\bibfnamefont{I.}~\bibnamefont{Van~Driessche}},
  \bibnamefont{and} \bibinfo{author}{\bibfnamefont{P.}~\bibnamefont{Bultinck}},
  \bibinfo{journal}{J. Am. Ceram. Soc.} pp.~\bibinfo{pages}{--}
  (\bibinfo{year}{2013}), \bibinfo{note}{accepted}.

\bibitem[{fn:({\natexlab{a}})}]{fn:Alio:CompChargeVacDef}
\bibinfo{note}{It is at this point important to note that all the systems under
  study in this work are charge neutral from the electronic point of view,
  \emph{i.e.} there are as many positive as negative charges in each cell. As
  such the term ``charge compensating vacancy'' and reference to it by the use
  of terms like ``charge compensation'' may be considered confusing. The
  terminology, however, originates in the study of ionic conductivity. There,
  the substitution of one cation by another will, when using the
  Kr{\"{o}}ger-Vink notation, always indicate the change in valence at the
  substitution site as a charge on the substituent element. For example, if
  \ce{MgO} is dissolved in \ce{CeO2}, tetravalent Ce is substituted by divalent
  Mg, which in Kr{\"{o}}ger-Vink notation can be written
  as:\cite{KrogerVink1956} \begin{center} \ce{MgO ->[\ce{CeO2}]
  Mg_{Ce}}$^{\prime\prime}$\ce{ + O_{O}}$^{\times}$\ce{ +
  V_{O}}$^{\bullet\bullet}$. \end{center} The right hand side shows how one
  \ce{CeO2} unit is replaced by a \ce{MgO} unit: The Mg at a Ce site, in which
  case the oxidation state of the site changes from IV to II, indicated as a
  double negative charge at the site; the O atom at an O site, with no change
  of the oxidation state; and a vacancy (V) at an oxygen site, changing the
  site's oxidation state from $-$II to $0$, indicated as a double positive
  charge. If within such a description the vacancy is omitted, the sum of the
  site charges will be non-zero. For this reason the vacancy is referred to as
  a ``charge compensating vacancy''. One may remark that one is actually
  balancing oxidation states (or valencies,
  \textit{cf.}~Ref.~\cite{fn:ValenceDisclaimer}) and not charges, and as such
  ``valence compensating vacancy'' would be a better suited term. However,
  charge compensating vacancy is the terminology used in (experimental)
  literature, so we will also use it to keep this link, and to avoid confusion
  on this account.}

\bibitem[{\citenamefont{Kundakovic and
  Flytzani-Stephanopoulos}(1998{\natexlab{a}})}]{KundakovicLj:ApplCatalA1998}
\bibinfo{author}{\bibfnamefont{L.}~\bibnamefont{Kundakovic}} \bibnamefont{and}
  \bibinfo{author}{\bibfnamefont{M.}~\bibnamefont{Flytzani-Stephanopoulos}},
  \bibinfo{journal}{Appl. Catal. A} \textbf{\bibinfo{volume}{171}},
  \bibinfo{pages}{13} (\bibinfo{year}{1998}{\natexlab{a}}), ISSN
  \bibinfo{issn}{0926-860X},
  \urlprefix\url{http://www.sciencedirect.com/science/article/pii/S0926860X98000568}.

\bibitem[{\citenamefont{Kundakovic and
  Flytzani-Stephanopoulos}(1998{\natexlab{b}})}]{KundakovicLj:JCat1998}
\bibinfo{author}{\bibfnamefont{L.}~\bibnamefont{Kundakovic}} \bibnamefont{and}
  \bibinfo{author}{\bibfnamefont{M.}~\bibnamefont{Flytzani-Stephanopoulos}},
  \bibinfo{journal}{J. Catal.} \textbf{\bibinfo{volume}{179}},
  \bibinfo{pages}{203 } (\bibinfo{year}{1998}{\natexlab{b}}).

\bibitem[{\citenamefont{Wang et~al.}(2005)\citenamefont{Wang, Rodriguez,
  Hanson, Gamarra, Mart\'{i}nez-Arias, and
  Fern\'{a}ndez-Garc\'{i}a}}]{WangX:JPhysChemB2005}
\bibinfo{author}{\bibfnamefont{X.}~\bibnamefont{Wang}},
  \bibinfo{author}{\bibfnamefont{J.~A.} \bibnamefont{Rodriguez}},
  \bibinfo{author}{\bibfnamefont{J.~C.} \bibnamefont{Hanson}},
  \bibinfo{author}{\bibfnamefont{D.}~\bibnamefont{Gamarra}},
  \bibinfo{author}{\bibfnamefont{A.}~\bibnamefont{Mart\'{i}nez-Arias}},
  \bibnamefont{and}
  \bibinfo{author}{\bibfnamefont{M.}~\bibnamefont{Fern\'{a}ndez-Garc\'{i}a}},
  \bibinfo{journal}{J. Phys. Chem. B} \textbf{\bibinfo{volume}{109}},
  \bibinfo{pages}{19595} (\bibinfo{year}{2005}),
  \eprint{http://pubs.acs.org/doi/pdf/10.1021/jp051970h},
  \urlprefix\url{http://pubs.acs.org/doi/abs/10.1021/jp051970h}.

\bibitem[{\citenamefont{Wang et~al.}(2006)\citenamefont{Wang, Rodriguez,
  Hanson, Gamarra, Mart\'{i}nez-Arias, and
  Fern\'{a}ndez-Garc\'{i}a}}]{WangX:JPhysChemB2006}
\bibinfo{author}{\bibfnamefont{X.}~\bibnamefont{Wang}},
  \bibinfo{author}{\bibfnamefont{J.~A.} \bibnamefont{Rodriguez}},
  \bibinfo{author}{\bibfnamefont{J.~C.} \bibnamefont{Hanson}},
  \bibinfo{author}{\bibfnamefont{D.}~\bibnamefont{Gamarra}},
  \bibinfo{author}{\bibfnamefont{A.}~\bibnamefont{Mart\'{i}nez-Arias}},
  \bibnamefont{and}
  \bibinfo{author}{\bibfnamefont{M.}~\bibnamefont{Fern\'{a}ndez-Garc\'{i}a}},
  \bibinfo{journal}{J. Phys. Chem. B} \textbf{\bibinfo{volume}{110}},
  \bibinfo{pages}{428} (\bibinfo{year}{2006}),
  \eprint{http://pubs.acs.org/doi/pdf/10.1021/jp055467g},
  \urlprefix\url{http://pubs.acs.org/doi/abs/10.1021/jp055467g}.

\bibitem[{\citenamefont{de~Biasi and Grillo}(2005)}]{deBiasi:JSolStateChem2005}
\bibinfo{author}{\bibfnamefont{R.}~\bibnamefont{de~Biasi}} \bibnamefont{and}
  \bibinfo{author}{\bibfnamefont{M.}~\bibnamefont{Grillo}},
  \bibinfo{journal}{J. Sol. State Chem.} \textbf{\bibinfo{volume}{178}},
  \bibinfo{pages}{1973} (\bibinfo{year}{2005}), ISSN \bibinfo{issn}{0022-4596}.

\bibitem[{\citenamefont{Li et~al.}(2010)\citenamefont{Li, Wei, and
  Pan}}]{LiB:IJHE2010}
\bibinfo{author}{\bibfnamefont{B.}~\bibnamefont{Li}},
  \bibinfo{author}{\bibfnamefont{X.}~\bibnamefont{Wei}}, \bibnamefont{and}
  \bibinfo{author}{\bibfnamefont{W.}~\bibnamefont{Pan}}, \bibinfo{journal}{Int.
  J. Hydrogen Energy} \textbf{\bibinfo{volume}{35}}, \bibinfo{pages}{3018 }
  (\bibinfo{year}{2010}).

\bibitem[{\citenamefont{Anwar et~al.}(2011)\citenamefont{Anwar, Kumar, Arshi,
  Ahmed, Seo, Lee, and Koo}}]{AnwarMS:JAlloysCompd2011}
\bibinfo{author}{\bibfnamefont{M.}~\bibnamefont{Anwar}},
  \bibinfo{author}{\bibfnamefont{S.}~\bibnamefont{Kumar}},
  \bibinfo{author}{\bibfnamefont{N.}~\bibnamefont{Arshi}},
  \bibinfo{author}{\bibfnamefont{F.}~\bibnamefont{Ahmed}},
  \bibinfo{author}{\bibfnamefont{Y.}~\bibnamefont{Seo}},
  \bibinfo{author}{\bibfnamefont{C.}~\bibnamefont{Lee}}, \bibnamefont{and}
  \bibinfo{author}{\bibfnamefont{B.~H.} \bibnamefont{Koo}},
  \bibinfo{journal}{J. Alloys Compd.} \textbf{\bibinfo{volume}{509}},
  \bibinfo{pages}{4525} (\bibinfo{year}{2011}), ISSN \bibinfo{issn}{0925-8388}.

\bibitem[{\citenamefont{Brisse and Knop}(1967)}]{BrisseF:CJChem1967}
\bibinfo{author}{\bibfnamefont{F.}~\bibnamefont{Brisse}} \bibnamefont{and}
  \bibinfo{author}{\bibfnamefont{O.}~\bibnamefont{Knop}},
  \bibinfo{journal}{Can. J. Chem.} \textbf{\bibinfo{volume}{45}},
  \bibinfo{pages}{609} (\bibinfo{year}{1967}).

\bibitem[{\citenamefont{Bae et~al.}(2004)\citenamefont{Bae, Choo, and
  Lee}}]{BaeJongSung:JoECS2004}
\bibinfo{author}{\bibfnamefont{J.~S.} \bibnamefont{Bae}},
  \bibinfo{author}{\bibfnamefont{W.~K.} \bibnamefont{Choo}}, \bibnamefont{and}
  \bibinfo{author}{\bibfnamefont{C.~H.} \bibnamefont{Lee}},
  \bibinfo{journal}{J. Eur. Ceram. Soc.} \textbf{\bibinfo{volume}{24}},
  \bibinfo{pages}{1291} (\bibinfo{year}{2004}).

\bibitem[{\citenamefont{de~Biasi and Grillo}(2008)}]{deBiasi:JAlloysCompd2008}
\bibinfo{author}{\bibfnamefont{R.}~\bibnamefont{de~Biasi}} \bibnamefont{and}
  \bibinfo{author}{\bibfnamefont{M.}~\bibnamefont{Grillo}},
  \bibinfo{journal}{J. Alloys Compd.} \textbf{\bibinfo{volume}{462}},
  \bibinfo{pages}{15} (\bibinfo{year}{2008}), ISSN \bibinfo{issn}{0925-8388}.

\bibitem[{\citenamefont{She et~al.}(2009)\citenamefont{She, Zheng, Li, Zhan,
  Chen, Zheng, and Lin}}]{SheYusheng:IJHE2009}
\bibinfo{author}{\bibfnamefont{Y.}~\bibnamefont{She}},
  \bibinfo{author}{\bibfnamefont{Q.}~\bibnamefont{Zheng}},
  \bibinfo{author}{\bibfnamefont{L.}~\bibnamefont{Li}},
  \bibinfo{author}{\bibfnamefont{Y.}~\bibnamefont{Zhan}},
  \bibinfo{author}{\bibfnamefont{C.}~\bibnamefont{Chen}},
  \bibinfo{author}{\bibfnamefont{Y.}~\bibnamefont{Zheng}}, \bibnamefont{and}
  \bibinfo{author}{\bibfnamefont{X.}~\bibnamefont{Lin}}, \bibinfo{journal}{Int.
  J. Hydrogen Energy} \textbf{\bibinfo{volume}{34}}, \bibinfo{pages}{8929}
  (\bibinfo{year}{2009}).

\bibitem[{\citenamefont{Ainirad et~al.}(2011)\citenamefont{Ainirad, Motlagh,
  and Maghsoudipoor}}]{AiniradA:JAlloysCompd2011}
\bibinfo{author}{\bibfnamefont{A.}~\bibnamefont{Ainirad}},
  \bibinfo{author}{\bibfnamefont{M.~K.} \bibnamefont{Motlagh}},
  \bibnamefont{and}
  \bibinfo{author}{\bibfnamefont{A.}~\bibnamefont{Maghsoudipoor}},
  \bibinfo{journal}{J. Alloys Compd.} \textbf{\bibinfo{volume}{509}},
  \bibinfo{pages}{1505} (\bibinfo{year}{2011}), ISSN \bibinfo{issn}{0925-8388}.

\bibitem[{\citenamefont{Tiwari et~al.}(2006)\citenamefont{Tiwari, Bhosle,
  Ramachandran, Sudhakar, Narayan, Budak, and
  Gupta}}]{TiwariA:ApplPhysLett2006}
\bibinfo{author}{\bibfnamefont{A.}~\bibnamefont{Tiwari}},
  \bibinfo{author}{\bibfnamefont{V.~M.} \bibnamefont{Bhosle}},
  \bibinfo{author}{\bibfnamefont{S.}~\bibnamefont{Ramachandran}},
  \bibinfo{author}{\bibfnamefont{N.}~\bibnamefont{Sudhakar}},
  \bibinfo{author}{\bibfnamefont{J.}~\bibnamefont{Narayan}},
  \bibinfo{author}{\bibfnamefont{S.}~\bibnamefont{Budak}}, \bibnamefont{and}
  \bibinfo{author}{\bibfnamefont{A.}~\bibnamefont{Gupta}},
  \bibinfo{journal}{Appl. Phys. Lett.} \textbf{\bibinfo{volume}{88}},
  \bibinfo{eid}{142511} (pages~\bibinfo{numpages}{3}) (\bibinfo{year}{2006}),
  \urlprefix\url{http://link.aip.org/link/?APL/88/142511/1}.

\bibitem[{\citenamefont{Vodungbo et~al.}(2007)\citenamefont{Vodungbo, Zheng,
  Vidal, Demaille, Etgens, and Mosca}}]{VodungboB:ApplyPhysLett2007}
\bibinfo{author}{\bibfnamefont{B.}~\bibnamefont{Vodungbo}},
  \bibinfo{author}{\bibfnamefont{Y.}~\bibnamefont{Zheng}},
  \bibinfo{author}{\bibfnamefont{F.}~\bibnamefont{Vidal}},
  \bibinfo{author}{\bibfnamefont{D.}~\bibnamefont{Demaille}},
  \bibinfo{author}{\bibfnamefont{V.~H.} \bibnamefont{Etgens}},
  \bibnamefont{and} \bibinfo{author}{\bibfnamefont{D.~H.} \bibnamefont{Mosca}},
  \bibinfo{journal}{Appl. Phys. Lett.} \textbf{\bibinfo{volume}{90}},
  \bibinfo{eid}{062510} (pages~\bibinfo{numpages}{3}) (\bibinfo{year}{2007}),
  \urlprefix\url{http://link.aip.org/link/?APL/90/062510/1}.

\bibitem[{\citenamefont{Fernandes et~al.}(2007)\citenamefont{Fernandes, Klein,
  Mattoso, Mosca, Silveira, Ribeiro, Schreiner, Varalda, and
  de~Oliveira}}]{FernandesV:PhysRevB2007}
\bibinfo{author}{\bibfnamefont{V.}~\bibnamefont{Fernandes}},
  \bibinfo{author}{\bibfnamefont{J.~J.} \bibnamefont{Klein}},
  \bibinfo{author}{\bibfnamefont{N.}~\bibnamefont{Mattoso}},
  \bibinfo{author}{\bibfnamefont{D.~H.} \bibnamefont{Mosca}},
  \bibinfo{author}{\bibfnamefont{E.}~\bibnamefont{Silveira}},
  \bibinfo{author}{\bibfnamefont{E.}~\bibnamefont{Ribeiro}},
  \bibinfo{author}{\bibfnamefont{W.~H.} \bibnamefont{Schreiner}},
  \bibinfo{author}{\bibfnamefont{J.}~\bibnamefont{Varalda}}, \bibnamefont{and}
  \bibinfo{author}{\bibfnamefont{A.~J.~A.} \bibnamefont{de~Oliveira}},
  \bibinfo{journal}{Phys. Rev. B} \textbf{\bibinfo{volume}{75}},
  \bibinfo{pages}{121304} (\bibinfo{year}{2007}),
  \urlprefix\url{http://link.aps.org/doi/10.1103/PhysRevB.75.121304}.

\bibitem[{\citenamefont{Song et~al.}(2007)\citenamefont{Song, Zhang, Wen, Zhu,
  and Xiao}}]{SongYQ:JApplPhys2007}
\bibinfo{author}{\bibfnamefont{Y.~Q.} \bibnamefont{Song}},
  \bibinfo{author}{\bibfnamefont{H.~W.} \bibnamefont{Zhang}},
  \bibinfo{author}{\bibfnamefont{Q.~Y.} \bibnamefont{Wen}},
  \bibinfo{author}{\bibfnamefont{H.}~\bibnamefont{Zhu}}, \bibnamefont{and}
  \bibinfo{author}{\bibfnamefont{J.~Q.} \bibnamefont{Xiao}},
  \bibinfo{journal}{J. Appl. Phys.} \textbf{\bibinfo{volume}{102}},
  \bibinfo{eid}{043912} (pages~\bibinfo{numpages}{5}) (\bibinfo{year}{2007}),
  \urlprefix\url{http://link.aip.org/link/?JAP/102/043912/1}.

\bibitem[{\citenamefont{Wen et~al.}(2007)\citenamefont{Wen, Zhang, Song, Yang,
  Zhu, and Xiao}}]{WenQY:JPhysCondensMatter2007}
\bibinfo{author}{\bibfnamefont{Q.-Y.} \bibnamefont{Wen}},
  \bibinfo{author}{\bibfnamefont{H.-W.} \bibnamefont{Zhang}},
  \bibinfo{author}{\bibfnamefont{Y.-Q.} \bibnamefont{Song}},
  \bibinfo{author}{\bibfnamefont{Q.-H.} \bibnamefont{Yang}},
  \bibinfo{author}{\bibfnamefont{H.}~\bibnamefont{Zhu}}, \bibnamefont{and}
  \bibinfo{author}{\bibfnamefont{J.~Q.} \bibnamefont{Xiao}},
  \bibinfo{journal}{J. Phys.: Condens. Matter} \textbf{\bibinfo{volume}{19}},
  \bibinfo{pages}{246205} (\bibinfo{year}{2007}),
  \urlprefix\url{http://stacks.iop.org/0953-8984/19/i=24/a=246205}.

\bibitem[{\citenamefont{Singhal et~al.}(2011)\citenamefont{Singhal, Kumari,
  Kumar, Dolia, Xing, Alzamora, Deshpande, Shripathi, and
  Saitovitch}}]{SinghalRK:JPhysDApplPhys2011}
\bibinfo{author}{\bibfnamefont{R.~K.} \bibnamefont{Singhal}},
  \bibinfo{author}{\bibfnamefont{P.}~\bibnamefont{Kumari}},
  \bibinfo{author}{\bibfnamefont{S.}~\bibnamefont{Kumar}},
  \bibinfo{author}{\bibfnamefont{S.~N.} \bibnamefont{Dolia}},
  \bibinfo{author}{\bibfnamefont{Y.~T.} \bibnamefont{Xing}},
  \bibinfo{author}{\bibfnamefont{M.}~\bibnamefont{Alzamora}},
  \bibinfo{author}{\bibfnamefont{U.~P.} \bibnamefont{Deshpande}},
  \bibinfo{author}{\bibfnamefont{T.}~\bibnamefont{Shripathi}},
  \bibnamefont{and}
  \bibinfo{author}{\bibfnamefont{E.}~\bibnamefont{Saitovitch}},
  \bibinfo{journal}{J. Phys. D: Appl. Phys.} \textbf{\bibinfo{volume}{44}},
  \bibinfo{pages}{165002} (\bibinfo{year}{2011}),
  \urlprefix\url{http://stacks.iop.org/0022-3727/44/i=16/a=165002}.

\bibitem[{\citenamefont{Sacanell et~al.}(2012)\citenamefont{Sacanell, Paulin,
  Ferrari, Garbarino, and Leyva}}]{SacanellJ:ApplPhysLett2012}
\bibinfo{author}{\bibfnamefont{J.}~\bibnamefont{Sacanell}},
  \bibinfo{author}{\bibfnamefont{M.~A.} \bibnamefont{Paulin}},
  \bibinfo{author}{\bibfnamefont{V.}~\bibnamefont{Ferrari}},
  \bibinfo{author}{\bibfnamefont{G.}~\bibnamefont{Garbarino}},
  \bibnamefont{and} \bibinfo{author}{\bibfnamefont{A.~G.} \bibnamefont{Leyva}},
  \bibinfo{journal}{Appl. Phys. Lett.} \textbf{\bibinfo{volume}{100}},
  \bibinfo{eid}{172405} (pages~\bibinfo{numpages}{3}) (\bibinfo{year}{2012}),
  \urlprefix\url{http://link.aip.org/link/?APL/100/172405/1}.

\bibitem[{\citenamefont{Xu et~al.}(2011)\citenamefont{Xu, Liu, Xu, Yan, Pei,
  Zhu, Wang, and Su}}]{XuD:SolStateIonics2011}
\bibinfo{author}{\bibfnamefont{D.}~\bibnamefont{Xu}},
  \bibinfo{author}{\bibfnamefont{X.}~\bibnamefont{Liu}},
  \bibinfo{author}{\bibfnamefont{S.}~\bibnamefont{Xu}},
  \bibinfo{author}{\bibfnamefont{D.}~\bibnamefont{Yan}},
  \bibinfo{author}{\bibfnamefont{L.}~\bibnamefont{Pei}},
  \bibinfo{author}{\bibfnamefont{C.}~\bibnamefont{Zhu}},
  \bibinfo{author}{\bibfnamefont{D.}~\bibnamefont{Wang}}, \bibnamefont{and}
  \bibinfo{author}{\bibfnamefont{W.}~\bibnamefont{Su}}, \bibinfo{journal}{Solid
  State Ionics} \textbf{\bibinfo{volume}{192}}, \bibinfo{pages}{510}
  (\bibinfo{year}{2011}), ISSN \bibinfo{issn}{0167-2738},
  \urlprefix\url{http://www.sciencedirect.com/science/article/pii/S016727381000144X}.

\bibitem[{\citenamefont{Yao et~al.}(2012)\citenamefont{Yao, Zhao, Chen, Wang,
  Ge, Wang, and Li}}]{YaoHC:JPowerSources2012}
\bibinfo{author}{\bibfnamefont{H.-C.} \bibnamefont{Yao}},
  \bibinfo{author}{\bibfnamefont{X.-L.} \bibnamefont{Zhao}},
  \bibinfo{author}{\bibfnamefont{X.}~\bibnamefont{Chen}},
  \bibinfo{author}{\bibfnamefont{J.-C.} \bibnamefont{Wang}},
  \bibinfo{author}{\bibfnamefont{Q.-Q.} \bibnamefont{Ge}},
  \bibinfo{author}{\bibfnamefont{J.-S.} \bibnamefont{Wang}}, \bibnamefont{and}
  \bibinfo{author}{\bibfnamefont{Z.-J.} \bibnamefont{Li}}, \bibinfo{journal}{J.
  Power Sources} \textbf{\bibinfo{volume}{205}}, \bibinfo{pages}{180}
  (\bibinfo{year}{2012}), ISSN \bibinfo{issn}{0378-7753},
  \urlprefix\url{http://www.sciencedirect.com/science/article/pii/S0378775312001759}.

\bibitem[{\citenamefont{Van~Horn}(2001)}]{Shannon:table}
\bibinfo{author}{\bibfnamefont{J.~D.} \bibnamefont{Van~Horn}},
  \emph{\bibinfo{title}{{Electronic Table of Shannon Ionic Radii}}}
  (\bibinfo{year}{2001}),
  \bibinfo{note}{http://v.web.umkc.edu/vanhornj/shannonradii.htm\ downloaded
  08/13/2010}.

\bibitem[{\citenamefont{Shannon}(1976)}]{Shannon:ACSA1976}
\bibinfo{author}{\bibfnamefont{R.~D.} \bibnamefont{Shannon}},
  \bibinfo{journal}{Acta Cryst.} \textbf{\bibinfo{volume}{A32}},
  \bibinfo{pages}{751} (\bibinfo{year}{1976}).

\bibitem[{\citenamefont{Bl{\"o}chl}(1994)}]{Blochl:prb94}
\bibinfo{author}{\bibfnamefont{P.~E.} \bibnamefont{Bl{\"o}chl}},
  \bibinfo{journal}{Phys. Rev. B} \textbf{\bibinfo{volume}{50}},
  \bibinfo{pages}{17953} (\bibinfo{year}{1994}).

\bibitem[{\citenamefont{Kresse and Joubert}(1999)}]{Kresse:prb99}
\bibinfo{author}{\bibfnamefont{G.}~\bibnamefont{Kresse}} \bibnamefont{and}
  \bibinfo{author}{\bibfnamefont{D.}~\bibnamefont{Joubert}},
  \bibinfo{journal}{Phys. Rev. B} \textbf{\bibinfo{volume}{59}},
  \bibinfo{pages}{1758} (\bibinfo{year}{1999}).

\bibitem[{\citenamefont{Ceperley and Alder}(1980)}]{CA:prl1980}
\bibinfo{author}{\bibfnamefont{D.~M.} \bibnamefont{Ceperley}} \bibnamefont{and}
  \bibinfo{author}{\bibfnamefont{B.~J.} \bibnamefont{Alder}},
  \bibinfo{journal}{Phys. Rev. Lett.} \textbf{\bibinfo{volume}{45}},
  \bibinfo{pages}{566} (\bibinfo{year}{1980}).

\bibitem[{\citenamefont{Perdew et~al.}(1996)\citenamefont{Perdew, Burke, and
  Ernzerhof}}]{PBE_1996prl}
\bibinfo{author}{\bibfnamefont{J.~P.} \bibnamefont{Perdew}},
  \bibinfo{author}{\bibfnamefont{K.}~\bibnamefont{Burke}}, \bibnamefont{and}
  \bibinfo{author}{\bibfnamefont{M.}~\bibnamefont{Ernzerhof}},
  \bibinfo{journal}{Phys. Rev. Lett.} \textbf{\bibinfo{volume}{77}},
  \bibinfo{pages}{3865} (\bibinfo{year}{1996}).

\bibitem[{\citenamefont{Kresse and Hafner}(1993)}]{Kresse:prb93}
\bibinfo{author}{\bibfnamefont{G.}~\bibnamefont{Kresse}} \bibnamefont{and}
  \bibinfo{author}{\bibfnamefont{J.}~\bibnamefont{Hafner}},
  \bibinfo{journal}{Phys. Rev. B} \textbf{\bibinfo{volume}{47}},
  \bibinfo{pages}{558} (\bibinfo{year}{1993}).

\bibitem[{\citenamefont{Kresse and Furthm\"uller}(1996)}]{Kresse:prb96}
\bibinfo{author}{\bibfnamefont{G.}~\bibnamefont{Kresse}} \bibnamefont{and}
  \bibinfo{author}{\bibfnamefont{J.}~\bibnamefont{Furthm\"uller}},
  \bibinfo{journal}{Phys. Rev. B} \textbf{\bibinfo{volume}{54}},
  \bibinfo{pages}{11169} (\bibinfo{year}{1996}).

\bibitem[{fn:({\natexlab{b}})}]{fn:missLDA}
\bibinfo{note}{LDA values for Sm, Gd, and Yb are missing since no LDA PAW
  potentials are available in the used distribution of the \textsc{vasp}
  program.}

\bibitem[{\citenamefont{Maradudin et~al.}(1971)\citenamefont{Maradudin,
  Montroll, Weiss, and Ipatova}}]{Maradudin:TheoryLattDynHarmApprox1971}
\bibinfo{author}{\bibfnamefont{A.~A.} \bibnamefont{Maradudin}},
  \bibinfo{author}{\bibfnamefont{E.~W.} \bibnamefont{Montroll}},
  \bibinfo{author}{\bibfnamefont{G.~H.} \bibnamefont{Weiss}}, \bibnamefont{and}
  \bibinfo{author}{\bibfnamefont{I.~P.} \bibnamefont{Ipatova}},
  \emph{\bibinfo{title}{Theory of lattice dynamics in the harmonic
  approximation}} (\bibinfo{publisher}{Academic press}, \bibinfo{address}{New
  York}, \bibinfo{year}{1971}), \bibinfo{edition}{2nd} ed.

\bibitem[{\citenamefont{Blanco et~al.}(1996)\citenamefont{Blanco, Pendas,
  Francisco, Recio, and Franco}}]{BlancoMA:JMolStruc1996}
\bibinfo{author}{\bibfnamefont{M.~A.} \bibnamefont{Blanco}},
  \bibinfo{author}{\bibfnamefont{A.~M.} \bibnamefont{Pendas}},
  \bibinfo{author}{\bibfnamefont{E.}~\bibnamefont{Francisco}},
  \bibinfo{author}{\bibfnamefont{J.~M.} \bibnamefont{Recio}}, \bibnamefont{and}
  \bibinfo{author}{\bibfnamefont{R.}~\bibnamefont{Franco}},
  \bibinfo{journal}{Theochem-J. Mol. Struct.} \textbf{\bibinfo{volume}{368}},
  \bibinfo{pages}{245} (\bibinfo{year}{1996}), ISSN \bibinfo{issn}{0166-1280}.

\bibitem[{\citenamefont{Francisco et~al.}(2001)\citenamefont{Francisco, Blanco,
  and Sanjurjo}}]{FranciscoE:PRB2001}
\bibinfo{author}{\bibfnamefont{E.}~\bibnamefont{Francisco}},
  \bibinfo{author}{\bibfnamefont{M.~A.} \bibnamefont{Blanco}},
  \bibnamefont{and} \bibinfo{author}{\bibfnamefont{G.}~\bibnamefont{Sanjurjo}},
  \bibinfo{journal}{Phys. Rev. B} \textbf{\bibinfo{volume}{63}},
  \bibinfo{pages}{094107} (\bibinfo{year}{2001}),
  \urlprefix\url{http://link.aps.org/doi/10.1103/PhysRevB.63.094107}.

\bibitem[{\citenamefont{Vanpoucke}(2011)}]{HIVE_REFERENCE}
\bibinfo{author}{\bibfnamefont{D.~E.~P.} \bibnamefont{Vanpoucke}},
  \emph{\bibinfo{title}{{HIVE} v2.1}} (\bibinfo{year}{2011}),
  \bibinfo{note}{{http://users.ugent.be/\~devpouck/hive\_refman/index.html}}.

\bibitem[{\citenamefont{Murnaghan}(1944)}]{MurnaghanFD:PNAS1944}
\bibinfo{author}{\bibfnamefont{F.~D.} \bibnamefont{Murnaghan}},
  \bibinfo{journal}{Proc. Natl. Acad. Sci. USA} \textbf{\bibinfo{volume}{30}},
  \bibinfo{pages}{244} (\bibinfo{year}{1944}).

\bibitem[{\citenamefont{Birch}(1947)}]{BirchF:PhysRev1947}
\bibinfo{author}{\bibfnamefont{F.}~\bibnamefont{Birch}},
  \bibinfo{journal}{Phys. Rev.} \textbf{\bibinfo{volume}{71}},
  \bibinfo{pages}{809} (\bibinfo{year}{1947}).

\bibitem[{\citenamefont{Monkhorst and Pack}(1976)}]{Monkhorst:prb76}
\bibinfo{author}{\bibfnamefont{H.~J.} \bibnamefont{Monkhorst}}
  \bibnamefont{and} \bibinfo{author}{\bibfnamefont{J.~D.} \bibnamefont{Pack}},
  \bibinfo{journal}{Phys. Rev. B} \textbf{\bibinfo{volume}{13}},
  \bibinfo{pages}{5188} (\bibinfo{year}{1976}).

\bibitem[{fn:({\natexlab{c}})}]{fn:lowConc}
\bibinfo{note}{For low oxygen vacancy concentrations, we assume that results
  for random distributions of vacancies can be approximated as linear
  combinations of configurations as the ones studied in this work. However, to
  retain a clear image of the specific influences different configurations
  have, we will only investigate these homogeneous distributions of vacancies.}

\bibitem[{\citenamefont{Deganello and Martorana}(2002)}]{DeganelloF:SSI2002}
\bibinfo{author}{\bibfnamefont{F.}~\bibnamefont{Deganello}} \bibnamefont{and}
  \bibinfo{author}{\bibfnamefont{A.}~\bibnamefont{Martorana}},
  \bibinfo{journal}{J. Solid State Chem.} \textbf{\bibinfo{volume}{163}},
  \bibinfo{pages}{527} (\bibinfo{year}{2002}).

\bibitem[{\citenamefont{Deganello et~al.}(2003)\citenamefont{Deganello, Longo,
  and Martorana}}]{DeganelloF:SSI2003}
\bibinfo{author}{\bibfnamefont{F.}~\bibnamefont{Deganello}},
  \bibinfo{author}{\bibfnamefont{A.}~\bibnamefont{Longo}}, \bibnamefont{and}
  \bibinfo{author}{\bibfnamefont{A.}~\bibnamefont{Martorana}},
  \bibinfo{journal}{J. Solid State Chem.} \textbf{\bibinfo{volume}{175}},
  \bibinfo{pages}{289 } (\bibinfo{year}{2003}), ISSN \bibinfo{issn}{0022-4596}.

\bibitem[{\citenamefont{Denton and
  Ashcroft}(1991)}]{VegardsLaw_DentonAR:PhysRevA1991}
\bibinfo{author}{\bibfnamefont{A.~R.} \bibnamefont{Denton}} \bibnamefont{and}
  \bibinfo{author}{\bibfnamefont{N.~W.} \bibnamefont{Ashcroft}},
  \bibinfo{journal}{Phys. Rev. A} \textbf{\bibinfo{volume}{43}},
  \bibinfo{pages}{3161} (\bibinfo{year}{1991}).

\bibitem[{\citenamefont{Morris et~al.}(1993)\citenamefont{Morris, Flavell,
  Mackrodt, and Morris}}]{MorrisBC:JMatChem1993}
\bibinfo{author}{\bibfnamefont{B.~C.} \bibnamefont{Morris}},
  \bibinfo{author}{\bibfnamefont{W.~R.} \bibnamefont{Flavell}},
  \bibinfo{author}{\bibfnamefont{W.~C.} \bibnamefont{Mackrodt}},
  \bibnamefont{and} \bibinfo{author}{\bibfnamefont{M.~A.}
  \bibnamefont{Morris}}, \bibinfo{journal}{J. Mater. Chem.}
  \textbf{\bibinfo{volume}{3}}, \bibinfo{pages}{1007} (\bibinfo{year}{1993}).

\bibitem[{\citenamefont{Belli{\`{e}}re
  et~al.}(2006)\citenamefont{Belli{\`{e}}re, Joorst, Stephan, de~Groot, and
  Weckhuysen}}]{BelliereV:JPhysChemB2006}
\bibinfo{author}{\bibfnamefont{V.}~\bibnamefont{Belli{\`{e}}re}},
  \bibinfo{author}{\bibfnamefont{G.}~\bibnamefont{Joorst}},
  \bibinfo{author}{\bibfnamefont{O.}~\bibnamefont{Stephan}},
  \bibinfo{author}{\bibfnamefont{F.~M.~F.} \bibnamefont{de~Groot}},
  \bibnamefont{and} \bibinfo{author}{\bibfnamefont{B.~M.}
  \bibnamefont{Weckhuysen}}, \bibinfo{journal}{J. Phys. Chem. B}
  \textbf{\bibinfo{volume}{110}}, \bibinfo{pages}{9984} (\bibinfo{year}{2006}).

\bibitem[{fn:({\natexlab{d}})}]{fn:ValenceDisclaimer}
\bibinfo{note}{In \textit{ab-initio} calculations as presented in this work,
  the oxidation state of the atoms is not strictly defined, and our reference
  to any type of oxidation state should not be taken as an absolute truth, but
  rather an educated guess. All elements used as dopants in this work have
  either a single (most common) oxidation state which is different from IV
  (\textit{e.g.} Zn) or are multivalent with most common oxidation states
  different from IV (\textit{e.g.} Yb or V). Since Shannon crystal radii are
  given both for different coordinations and different oxidation states (though
  some combinations which might be of interest for this work are missing) we
  have attempted to derive the oxidation state of the dopants in the presented
  systems, based on the magnetization of the ground state system, under the
  assumption of integer values for this magnetization. In most cases these
  results pointed at the most common oxidation state, and in some cases
  degeneracies were present, with Co being the most extreme case
  (\textit{cf}.~Ref.~\cite{fn:magCo}). In cases where the oxidation state was
  uncertain, Shannon crystal radii in the ball-park of our calculated atomic
  radii were used as indicator for the dopant oxidation state. As a result, the
  stated oxidation numbers should only be considered as a guess, although they
  might point to underlying physical relations with the atomic oxidation
  states.}

\bibitem[{fn:({\natexlab{e}})}]{fn:magCo}
\bibinfo{note}{Of all systems investigated, Co doping is the most problematic
  one due to the near degeneracy of different magnetic configurations
  (magnetization varying from $1$ to $5\ \mu_{B}$ show differences in total
  energy of $\sim0.20$ and $\sim0.08$ eV for $3$\% doped systems).
  Interestingly enough, experiments seem to encounter similar variation in the
  observed magnetic moment, with values varying with the Co concentration,
  substrate, and deposition method. Where Tiwari \textit{et al.} present
  $6\mu_{B}$ at $3$\% Co doping, Vodungbo \textit{et al.} measure about
  $1.5\mu_{B}$ at $4.5$\%, while Fernandez \textit{et al.} and Song \textit{et
  al.} measure about $5\mu_{B}$ at concentrations of $5$ and $3$\% ,
  respectively.\cite{TiwariA:ApplPhysLett2006, VodungboB:ApplyPhysLett2007,
  FernandesV:PhysRevB2007, SongYQ:JApplPhys2007}}.

\bibitem[{\citenamefont{Cordero et~al.}(2008)\citenamefont{Cordero, Gomez,
  Platero-Prats, Reves, Echeverria, Cremades, Barragan, and
  Alvarez}}]{Cordero:DT2008}
\bibinfo{author}{\bibfnamefont{B.}~\bibnamefont{Cordero}},
  \bibinfo{author}{\bibfnamefont{V.}~\bibnamefont{Gomez}},
  \bibinfo{author}{\bibfnamefont{A.~E.} \bibnamefont{Platero-Prats}},
  \bibinfo{author}{\bibfnamefont{M.}~\bibnamefont{Reves}},
  \bibinfo{author}{\bibfnamefont{J.}~\bibnamefont{Echeverria}},
  \bibinfo{author}{\bibfnamefont{E.}~\bibnamefont{Cremades}},
  \bibinfo{author}{\bibfnamefont{F.}~\bibnamefont{Barragan}}, \bibnamefont{and}
  \bibinfo{author}{\bibfnamefont{S.}~\bibnamefont{Alvarez}},
  \bibinfo{journal}{Dalton Transactions} pp. \bibinfo{pages}{2832--2838}
  (\bibinfo{year}{2008}), \urlprefix\url{http://dx.doi.org/10.1039/b801115j}.

\bibitem[{fn:({\natexlab{f}})}]{fn:DFTU_Sm}
\bibinfo{note}{The DFT+U study of Sm doped \ce{CeO2} by Ismail \textit{et al.}
  may be an indication that this also holds for PBE+U calculations, though they
  only included the Coulomb correction for
  Ce.\cite{IsmailA:PhysChemChemPhys2011}}.

\bibitem[{\citenamefont{Lin et~al.}(2003)\citenamefont{Lin, Luo, Zhong, Yan,
  Liu, and Liu}}]{LinR:ApplCatal2003}
\bibinfo{author}{\bibfnamefont{R.}~\bibnamefont{Lin}},
  \bibinfo{author}{\bibfnamefont{M.-F.} \bibnamefont{Luo}},
  \bibinfo{author}{\bibfnamefont{Y.-J.} \bibnamefont{Zhong}},
  \bibinfo{author}{\bibfnamefont{Z.-L.} \bibnamefont{Yan}},
  \bibinfo{author}{\bibfnamefont{G.-Y.} \bibnamefont{Liu}}, \bibnamefont{and}
  \bibinfo{author}{\bibfnamefont{W.-P.} \bibnamefont{Liu}},
  \bibinfo{journal}{Appl. Catal., A} \textbf{\bibinfo{volume}{255}},
  \bibinfo{pages}{331} (\bibinfo{year}{2003}), ISSN \bibinfo{issn}{0926-860X}.

\bibitem[{\citenamefont{Murgida et~al.}(2012)\citenamefont{Murgida, Vildosola,
  Ferrari, and Llois}}]{MurgidaGE:SolidStateComm2012}
\bibinfo{author}{\bibfnamefont{G.}~\bibnamefont{Murgida}},
  \bibinfo{author}{\bibfnamefont{V.}~\bibnamefont{Vildosola}},
  \bibinfo{author}{\bibfnamefont{V.}~\bibnamefont{Ferrari}}, \bibnamefont{and}
  \bibinfo{author}{\bibfnamefont{A.}~\bibnamefont{Llois}},
  \bibinfo{journal}{Solid State Communications} \textbf{\bibinfo{volume}{152}},
  \bibinfo{pages}{368} (\bibinfo{year}{2012}), ISSN \bibinfo{issn}{0038-1098},
  \urlprefix\url{http://www.sciencedirect.com/science/article/pii/S0038109811006806}.

\bibitem[{\citenamefont{Van~de Velde et~al.}(2012)\citenamefont{Van~de Velde,
  Bruggeman, Stove, Pollefeyt, Brunkahl, and
  Van~Driessche}}]{VandeVeldeNigel:EurJInorChem2012}
\bibinfo{author}{\bibfnamefont{N.}~\bibnamefont{Van~de Velde}},
  \bibinfo{author}{\bibfnamefont{T.}~\bibnamefont{Bruggeman}},
  \bibinfo{author}{\bibfnamefont{L.}~\bibnamefont{Stove}},
  \bibinfo{author}{\bibfnamefont{G.}~\bibnamefont{Pollefeyt}},
  \bibinfo{author}{\bibfnamefont{O.}~\bibnamefont{Brunkahl}}, \bibnamefont{and}
  \bibinfo{author}{\bibfnamefont{I.}~\bibnamefont{Van~Driessche}},
  \bibinfo{journal}{Eur. J. Inor. Chem.} \textbf{\bibinfo{volume}{2012}},
  \bibinfo{pages}{1186} (\bibinfo{year}{2012}), ISSN \bibinfo{issn}{1099-0682},
  \urlprefix\url{http://dx.doi.org/10.1002/ejic.201100951}.

\bibitem[{\citenamefont{Yamamura et~al.}(2004)\citenamefont{Yamamura, Nishino,
  and Kakinuma}}]{YamamuraH:JCSJ2004}
\bibinfo{author}{\bibfnamefont{H.}~\bibnamefont{Yamamura}},
  \bibinfo{author}{\bibfnamefont{H.}~\bibnamefont{Nishino}}, \bibnamefont{and}
  \bibinfo{author}{\bibfnamefont{K.}~\bibnamefont{Kakinuma}},
  \bibinfo{journal}{J. Ceram. Soc. Jpn} \textbf{\bibinfo{volume}{112}},
  \bibinfo{pages}{553} (\bibinfo{year}{2004}).

\bibitem[{\citenamefont{Gerward et~al.}(2005)\citenamefont{Gerward, Olsen,
  Petit, Vaitheeswaran, Kanchana, and Svane}}]{GerwardL:JAlloysCompd2005}
\bibinfo{author}{\bibfnamefont{L.}~\bibnamefont{Gerward}},
  \bibinfo{author}{\bibfnamefont{J.~S.} \bibnamefont{Olsen}},
  \bibinfo{author}{\bibfnamefont{L.}~\bibnamefont{Petit}},
  \bibinfo{author}{\bibfnamefont{G.}~\bibnamefont{Vaitheeswaran}},
  \bibinfo{author}{\bibfnamefont{V.}~\bibnamefont{Kanchana}}, \bibnamefont{and}
  \bibinfo{author}{\bibfnamefont{A.}~\bibnamefont{Svane}}, \bibinfo{journal}{J.
  Alloys Compd.} \textbf{\bibinfo{volume}{400}}, \bibinfo{pages}{56}
  (\bibinfo{year}{2005}), ISSN \bibinfo{issn}{0925-8388},
  \urlprefix\url{http://www.sciencedirect.com/science/article/pii/S0925838805003403}.

\bibitem[{\citenamefont{Duclos et~al.}(1988)\citenamefont{Duclos, Vohra, Ruoff,
  Jayaraman, and Espinosa}}]{DuclosSJ:RhysRevB1988}
\bibinfo{author}{\bibfnamefont{S.~J.} \bibnamefont{Duclos}},
  \bibinfo{author}{\bibfnamefont{Y.~K.} \bibnamefont{Vohra}},
  \bibinfo{author}{\bibfnamefont{A.~L.} \bibnamefont{Ruoff}},
  \bibinfo{author}{\bibfnamefont{A.}~\bibnamefont{Jayaraman}},
  \bibnamefont{and} \bibinfo{author}{\bibfnamefont{G.~P.}
  \bibnamefont{Espinosa}}, \bibinfo{journal}{Phys. Rev. B}
  \textbf{\bibinfo{volume}{38}}, \bibinfo{pages}{7755} (\bibinfo{year}{1988}),
  \urlprefix\url{http://link.aps.org/doi/10.1103/PhysRevB.38.7755}.

\bibitem[{\citenamefont{Nakajima et~al.}(1994)\citenamefont{Nakajima,
  Yoshihara, and Ishigame}}]{NakajimaA:PhysRevB1994}
\bibinfo{author}{\bibfnamefont{A.}~\bibnamefont{Nakajima}},
  \bibinfo{author}{\bibfnamefont{A.}~\bibnamefont{Yoshihara}},
  \bibnamefont{and} \bibinfo{author}{\bibfnamefont{M.}~\bibnamefont{Ishigame}},
  \bibinfo{journal}{Phys. Rev. B} \textbf{\bibinfo{volume}{50}},
  \bibinfo{pages}{13297} (\bibinfo{year}{1994}), \bibinfo{note}{bulk modulus
  calculated from elastic constants.},
  \urlprefix\url{http://link.aps.org/doi/10.1103/PhysRevB.50.13297}.

\bibitem[{\citenamefont{Tsuru et~al.}(2010)\citenamefont{Tsuru, Shinzato,
  Saito, Shimazu, Shiono, and Morinaga}}]{TsuruY:JCeramJpn2010}
\bibinfo{author}{\bibfnamefont{Y.}~\bibnamefont{Tsuru}},
  \bibinfo{author}{\bibfnamefont{Y.}~\bibnamefont{Shinzato}},
  \bibinfo{author}{\bibfnamefont{Y.}~\bibnamefont{Saito}},
  \bibinfo{author}{\bibfnamefont{M.}~\bibnamefont{Shimazu}},
  \bibinfo{author}{\bibfnamefont{M.}~\bibnamefont{Shiono}}, \bibnamefont{and}
  \bibinfo{author}{\bibfnamefont{M.}~\bibnamefont{Morinaga}},
  \bibinfo{journal}{J. Ceram. Soc. Jpn.} \textbf{\bibinfo{volume}{118}},
  \bibinfo{pages}{241} (\bibinfo{year}{2010}).

\bibitem[{\citenamefont{Fagg et~al.}(2006)\citenamefont{Fagg, Frade, Kharton,
  and Marozau}}]{FaggDP:JSolStateChem2006}
\bibinfo{author}{\bibfnamefont{D.}~\bibnamefont{Fagg}},
  \bibinfo{author}{\bibfnamefont{J.}~\bibnamefont{Frade}},
  \bibinfo{author}{\bibfnamefont{V.}~\bibnamefont{Kharton}}, \bibnamefont{and}
  \bibinfo{author}{\bibfnamefont{I.}~\bibnamefont{Marozau}},
  \bibinfo{journal}{J. Sol. State Chem.} \textbf{\bibinfo{volume}{179}},
  \bibinfo{pages}{1469} (\bibinfo{year}{2006}), ISSN \bibinfo{issn}{0022-4596}.

\bibitem[{fn:({\natexlab{g}})}]{fn:linTEC}
\bibinfo{note}{Note that the `linear' in linear TEC refers to thermal expansion
  in one dimension, and does not indicate any linearity with regard to this
  coefficient. Adding to the confusion however is the fact that for a large
  experimental temperature range the coefficient changes roughly linearly, as
  is shown in Fig.~\ref{fig:TECplotNV}.}

\bibitem[{fn:({\natexlab{h}})}]{fn:VacNrDef}
\bibinfo{note}{In this, $N_{vac}$ is related to $y$ via the relation
  $N_{vac}=yN_{uc}$ with $N_{uc}$ the number of unit cells required to build
  the supercell of the doped system.}

\bibitem[{\citenamefont{Catlow}(1983)}]{CatlowCRA:SolidStateIonics1983}
\bibinfo{author}{\bibfnamefont{C.}~\bibnamefont{Catlow}},
  \bibinfo{journal}{Solid State Ionics} \textbf{\bibinfo{volume}{8}},
  \bibinfo{pages}{89} (\bibinfo{year}{1983}), ISSN \bibinfo{issn}{0167-2738}.

\bibitem[{\citenamefont{Butler et~al.}(1983)\citenamefont{Butler, Catlow,
  Fender, and Harding}}]{ButlerV:SolStateIonics1983}
\bibinfo{author}{\bibfnamefont{V.}~\bibnamefont{Butler}},
  \bibinfo{author}{\bibfnamefont{C.}~\bibnamefont{Catlow}},
  \bibinfo{author}{\bibfnamefont{B.}~\bibnamefont{Fender}}, \bibnamefont{and}
  \bibinfo{author}{\bibfnamefont{J.}~\bibnamefont{Harding}},
  \bibinfo{journal}{Solid State Ionics} \textbf{\bibinfo{volume}{8}},
  \bibinfo{pages}{109} (\bibinfo{year}{1983}), ISSN \bibinfo{issn}{0167-2738}.

\bibitem[{\citenamefont{Kr\"{o}ger and Vink}(1956)}]{KrogerVink1956}
\bibinfo{author}{\bibfnamefont{F.~A.} \bibnamefont{Kr\"{o}ger}}
  \bibnamefont{and} \bibinfo{author}{\bibfnamefont{H.~J.} \bibnamefont{Vink}},
  \emph{\bibinfo{title}{Solid State Physics}}, vol.~\bibinfo{volume}{3}
  (\bibinfo{year}{1956}), \bibinfo{edition}{7th} ed., \bibinfo{note}{p.
  273--301}.

\bibitem[{\citenamefont{Ismail et~al.}(2011)\citenamefont{Ismail, Hooper,
  Giorgi, and Woo}}]{IsmailA:PhysChemChemPhys2011}
\bibinfo{author}{\bibfnamefont{A.}~\bibnamefont{Ismail}},
  \bibinfo{author}{\bibfnamefont{J.}~\bibnamefont{Hooper}},
  \bibinfo{author}{\bibfnamefont{J.~B.} \bibnamefont{Giorgi}},
  \bibnamefont{and} \bibinfo{author}{\bibfnamefont{T.~K.} \bibnamefont{Woo}},
  \bibinfo{journal}{Phys. Chem. Chem. Phys.} \textbf{\bibinfo{volume}{13}},
  \bibinfo{pages}{6116} (\bibinfo{year}{2011}),
  \urlprefix\url{http://dx.doi.org/10.1039/C0CP02062A}.

\end{thebibliography}

\end{document}